\documentclass[aps,twocolumn,amsmath,amssymb,floatfix,showpacs]{revtex4}
\pdfoutput=1

\usepackage{graphicx}
\usepackage{dcolumn} 
\usepackage{bm}      
\usepackage[usenames,dvipsnames]{color}
\usepackage{ulem} 
\begin{document}

\title{Anisotropy and non-universality {in scaling laws of the} large scale energy spectrum in rotating turbulence}
\author{Amrik Sen$^{1,2}$, Pablo D. Mininni$^{1,3}$, Duane Rosenberg$^1$ and Annick Pouquet$^1$.}
\affiliation{$^1$Institute for Mathematics Applied to Geosciences (IMAGe), 
         CISL, NCAR, P.O. Box 3000, Boulder, Colorado 80307-3000, USA. \\
  	     $^2$Department of Applied Mathematics, 526 UCB, University of 
         Colorado, Boulder, Colorado 80309-0526, USA.\\
             $^3$Departamento de F\'\i sica, Facultad de Ciencias Exactas y
         Naturales, Universidad de Buenos Aires and IFIBA, CONICET, Ciudad 
         Universitaria, 1428 Buenos Aires, Argentina.}
\date{\today}

\begin{abstract}
Rapidly rotating turbulent flow is characterized by the emergence of columnar structures that are representative of quasi-two dimensional behavior of the flow. It is known that when energy is injected into the fluid at an intermediate scale $L_f$, it cascades towards smaller as well as larger scales. In this paper we analyze the flow in the \textit{inverse cascade} range at a small but fixed Rossby number, {$\mathcal{R}o_f \approx 0.05$}. Several {numerical simulations with} helical and non-helical forcing functions are considered in periodic boxes with unit aspect ratio. In order to resolve the inverse cascade range with {reasonably} large Reynolds number, the analysis is based on large eddy simulations which include the effect of helicity on eddy viscosity and eddy noise. Thus, we model the small scales and resolve explicitly the large scales. We show that the large-scale energy spectrum has at least two solutions: one that is consistent with Kolmogorov-Kraichnan-Batchelor-Leith phenomenology for the inverse cascade of energy in two-dimensional (2D) turbulence with a {$\sim k_{\perp}^{-5/3}$} scaling, and the other that corresponds to a steeper {$\sim k_{\perp}^{-3}$} spectrum in which the three-dimensional (3D) modes release a substantial fraction of their energy per unit time to the 2D modes. {The spectrum that} emerges {depends on} the anisotropy of the forcing function{,} the former solution prevailing for forcings in which more energy is injected into the 2D modes while the latter prevails for isotropic forcing. {In the case of anisotropic forcing, whence the energy} goes from the 2D to the 3D modes at low wavenumbers, large-scale shear is created resulting in another time scale $\tau_{sh}$, associated with shear, {thereby producing} a $\sim k^{-1}$ spectrum for the {total energy} with the 2D modes still following a {$\sim k_{\perp}^{-5/3}$} scaling. 
\end{abstract}
\pacs{47.32.Ef, 47.27.Gs, 47.27.Jv}
\maketitle

\section{Introduction}

The emergence of anisotropy in rotating flows has been studied extensively since the seminal work of Taylor and Proudman \cite{Hough1897, Pdman16, Taylor17}. More recently, it has been observed both experimentally \cite{Moisy05, Lamriben11} and numerically \cite{Mininni10} for the velocity field, as well as for passive scalar fluctuations using a reduced model \cite{Sprague06, Grooms10} and direct numerical simulations (DNS) \cite{Imazio11}. While linear theory like the \textit{Taylor-Proudman} theorem (see, e.g., \cite{GSpan68}) explains the existence of the columnar structures in the laminar regime, more recent theories account for the nonlinearities in the flow and shed light on the mechanism of two dimensionalization of the flow from an initial isotropic state \cite{Cambon01, DSon06, Mahalov96}.

An interesting feature of such flows is the transfer of energy to large scales on application of external forcing \cite{Smith96, Smith99,Chen05}. However, there has been little consensus about the scaling of the energy spectrum at large scales. The numerical simulation of \cite{Smith99} reports a steep {$\sim k_{\perp}^{-3}$} spectrum whereas \cite{Smith96, Chen05} show a more conventional {$\sim k_{\perp}^{-5/3}$} scaling that is reminiscent of two-dimensional (2D) Kolmogorov-Kraichnan-Batchelor-Leith (KKBL) phenomenology for an inverse cascade of energy \cite{Kraichnan67} (also see \cite{Davidson04}). In \cite{Smith05}, a model is used to {show} that a {$\sim k_{\perp}^{-5/3}$} spectrum for the 2D modes (also called ``slow modes'') results when triadic interactions between the 2D and the 3D (``fast'') modes are {discounted for artificially}, but a {$\sim k_{\perp}^{-3}$} spectrum {is observed} when all interactions are {accounted for}. Besides, a $\sim k_{\perp}^{-3}$ law for the horizontal kinetic energy spectra is also observed in a rapidly rotating Rayleigh-B\'{e}nard convection using a reduced model \cite{Julien12}. It must be pointed out that the conserved quantities (for an inviscid fluid) in the case of two dimensional flows, viz., energy and enstrophy, are different from the three dimensional case, where energy and helicity {(the correlation between velocity and vorticity)} are conserved. Therefore,{ the physical mechanism leading to an inverse cascade of energy in the three dimensional case does not follow immediately from its two dimensional counterpart.} Besides, the nature of forcing, i.e. spectrally isotropic vs.~anisotropic (equivalently 3D vs.~2D), with three or two spatial components (3C vs.~2C), and the aspect ratio of the computational box may play significant roles in the dynamics of the flow.

In this paper, we revisit the issue of inverse cascade of energy in rotating flows within a specific framework, viz., fixed Rossby number and unit aspect ratio of the computational box. However, we explore different forcing functions to consider the effects of spectral anisotropy and of helicity in the inverse cascade range. We present results from numerical simulations that use a subgrid scale model developed in \cite{Baerenzung08}; this model was validated against DNS of rotating flows in \cite{Baeren10, Baeren11, trieste_12}. We show that the two observed spectra, viz., {$\sim k_{\perp}^{-5/3}$} and {$\sim k_{\perp}^{-3}$} can arise in full simulations (simulations that resolve all triadic interactions and account for coupling between the 2D and the 3D modes). When the forcing is isotropic, energy goes from the 3D to the 2D modes and a $\sim k_\perp^{-3}$ spectrum results for the energy in the slow modes. When more energy is pumped into the 2D modes, less energy goes from the 3D to the 2D modes and a {$\sim k_{\perp}^{-5/3}$} spectrum {is seen} for the slow modes.

The kinematics of the nonlinear advection term also changes significantly as the spectra of slow modes change from {$\sim k_{\perp}^{-3}$} to {$\sim k_{\perp}^{-5/3}$}. We study the velocity gradient tensor in all simulations and compute the largest eigenvalue of the rate of strain tensor. For the case of anisotropic forcing when the flux of energy between the 3D and the 2D modes reverses and energy at large scales goes from the 2D to the 3D modes, a {significant amount of} shear is created at large scales. {This introduces} a new \textit{shear timescale} $\tau_{sh}$ that is independent of wavenumber. As a result, the spectrum for the total energy approaches a $\sim k^{-1}$ power law.

The remainder of this paper is organized as follows. In Sec.~\ref{sec:waves} we discuss previous results, introduce equations and notations used in the rest of the paper, and derive equations to study the coupling between modes and the energy transfer between scales. In Sec.~\ref{sec:methods} we present the LES model used in the numerical simulations and describe all the runs as well as the different spectra used to characterize scaling laws in the inverse cascade range. Finally, in Sec.~\ref{sec:results} we present and discuss the numerical results, while in Sec.~\ref{sec:conclusion} we conclude with brief remarks and pointers to some open questions.

\section{Inertial waves and energy transfer to the slow manifold\label{sec:waves}}

\subsection{Equations}

The non-dimensionalized incompressible Navier-Stokes equations with global rotation, ${\bf \Omega}=\Omega\hat{z}$, are as follows:

\begin{align}
 \partial_t {\bf u} + ({\bf u}\cdot \nabla){\bf u} + \frac{1}{\mathcal{R}o}\hat{z}\times {\bf u} 
 &= -\nabla{P} + \frac{1}{\mathcal{R}e}\nabla^{2}{\bf u} + {\bf f}, \label{NSErot1} \\
 \nabla \cdot {\bf u} &= 0  \ , \label{incompress}
\end{align}
where ${\bf u}$ is the instantaneous velocity field, $P$ is the pressure term, ${\bf f}$ is an external force per unit of mass, the Rossby number is $\mathcal{R}o = U_0 (2 L_0 \Omega)^{-1}$ (where $U_0$ and $L_0$ are respectively normalized velocity and length scales taken to be unity) and the Reynolds number is $\mathcal{R}e = U_0L_0/\nu$ ({where} $\nu$ is the kinematic viscosity).

The forcing term ${\bf f}$ is introduced in the Navier-Stokes equation to study the inverse cascade of energy. {In the} simulations {presented in this paper}, besides $\mathcal{R}e$ and $\mathcal{R}o$ defined at characteristic length scales, we will be interested primarily in the Reynolds and Rossby numbers based on the forcing scale $L_f$, {at} which the external force is applied. {The latter quantities are defined as follows:}
\begin{equation}
\mathcal{R}e_f = \frac{L_f U}{\nu} ,
\end{equation}
and
\begin{equation}
\mathcal{R}o_f = \frac{U}{2 L_f \Omega} ,
\end{equation}
where $U$ is the r.m.s.~velocity before the inverse cascade {is initiated} (or equivalently, the r.m.s.~velocity at the forcing scale at any time {during} the simulation) in units of $U_0$. The time-scale associated with forcing wavenumber is defined as,
\begin{equation}
 \tau_f := \frac{L_f}{U}. \label{Tsc_f}
\end{equation}

\subsection{Resonant interactions, slow manifold, and large-scale structures} \label{ss:slow}

The linear, inviscid approximation of Eq.~\eqref{NSErot1} in the absence of forcing has wave solutions called inertial waves \cite{GSpan68}. In the language of \text{wave turbulence} theory, the inviscid version of Eq.~\eqref{NSErot1} can be re-written as \cite{Wal92}:
\begin{align}
\partial_t a^{s_{k}}(t) &= \mathcal{R}o \sum_{s_p, s_q}\int_{{\bf k} + {\bf p} + {\bf q} = 0}C_{kpq}^{s_k s_p s_q} a^{s_{p}^{\star}} a^{s_{q}^{\star}} \nonumber \\ 
 & \times e^{i(\omega_{s_k} + \omega_{s_p} + \omega_{s_q})t} dp dq,\label{NSErot2}
\end{align}
where $\star$ denotes complex conjugate, $C_{kpq}^{s_k s_p s_q} = (s_q q - s_p p)(h_{s_p}^* \times h_{s_q}^*) \cdot h_{s_k}^*/2$ is the modal transfer coefficient, and $\mathcal{R}o$ is assumed to be small and therefore represents rapid rotation (hence weak nonlinearity). Note that Eq.~\eqref{NSErot2} is not closed and, hence, any practical solution can be realized based on an equivalent closed set of equations (see e.g. \cite{Lesieur08} for a thorough discussion on closed models). Furthermore, the \textit{Craya-Herring} helical basis $h_s$ \cite{craya, herring} has been used in deriving Eq.~\eqref{NSErot2} with the canonical basis corresponding to a given wavevector ${\hat { k}}$, $\hat{\psi} := {\hat { k}} \times {\hat { z}}$ and $\hat {k} \times \hat{\psi}$. The amplitude $a^s$ of ${\bf u}$ is associated with the helical wave with a dispersion relation for the wave frequency: $\omega_s(k)=2\Omega s \frac{k_{\parallel}}{k}$. Each wave vector is associated with two waves of opposite polarization, $s=\pm 1$. Clearly, $\omega_s(k) = 0$ implies a flow restricted to a plane perpendicular to the rotation axis (i.e., $k_{\parallel}=0$). Hence, 2D modes are also known as \textit{slow} modes. In other words, 2D modes correspond to vortical motions with no fast wave modulation.

The mechanism of transfer of energy towards two dimensional modes {that is} responsible for the formation of Taylor columns is based on near resonant condition of the interacting triads \cite{Wal92}: 
\begin{equation}
 s_k\frac{k_{||}}{k} + s_p\frac{p_{||}}{p} + s_q\frac{q_{||}}{q}= \mathcal{O}(\mathcal{R}o) \text{ with } {\bf k}+{\bf p}+{\bf q} = 0 .
\label{resonant3}
\end{equation}
However, the problem with wave turbulence theory is that it is not valid for too small values of $k_{\parallel}$. In fact, the predicted energy transfer is zero for $k_{\parallel}=0$ \cite{Galtier03} {because} 2D and 3D modes are decoupled in such theories at lowest order. Similar analysis is presented using two-point closures of turbulence, such as the Eddy Damped Quasi-Normal Markovian (EDQNM) closure developed earlier in the context of rotating flows (see, e.g., \cite{cambon_89}). Even a sophisticated asymptotic quasi-normal Markovian theory, built on the EDQNM closure \cite{Bellet06, Sagaut08}, does not deal with $k_{\parallel}=0$. Thus, while the gradual concentration of energy in close proximity of the slow manifold can be theoretically justified to explain numerical and experimental observations, the exact coupling between the slow manifold and the 3D modes leading to a transfer of energy from 3D to 2D modes still remains an unresolved problem. The inverse cascade of energy, that will be further elaborated upon in Sec.~\ref{sec:results}, presumably happens in this slow manifold. 

An alternative theory on the egression of columnar structures is given by \cite{DSon06}; it is based on the conservation of linear momentum $P_z =\frac{1}{2} \int_{V_R} ({\bf x} \times {\bf \omega})_z \, dV$ and of angular momentum $L_z = \int_{V_R} ({\bf x} \times {\bf u})_zdV$ in the axial direction (within a cylinder of radius $R$), resulting in a relative concentration of the kinetic energy density within this cylinder where it disperses to form columnar clouds. This holds in the linear time scale $\Omega^{-1}$, when the non-linear term is small and hence can be neglected in comparison with the Coriolis term ($U_0 \ll \Omega L_0$). However, the percentage of total energy contained within the cylinder falls as $(\Omega t)^{-1}$, so the columns eventually become weak, although the energy density remains higher within the cylinder than outside. The time scale associated with this process, $\tau_{\Omega}\sim \Omega^{-1}$, will be {relevant for} the analysis of the inverse cascade regime in the following sections.

\subsection{Coupling between modes and energy transfer}

Lewis Fry Richardson's famous couplet, \lq \textit{Big whirls have little whirls that feed on their velocity, and little whirls have lesser whirls and so on to viscosity}\rq ~is the antithesis of observations made by experimentalists \cite{Moisy05} and analysts \cite{Smith99, Chen05} in the context of rotating turbulence. 

The notion of inverse cascade of energy to large scales is well known in 2D turbulence \cite{KraichMont80} (also see e.g., \cite{Davidson04}) and may be justified in simple terms on the basis of \textit{Fj{\o}rtoft's} theorem due to the conservation of quadratic invariants (see, e.g., \cite{Lesieur08}). In other words, nonlinear triadic interactions conserve both the energy and the enstrophy, $ Z:=\left< \omega^2 \right>/2$, and as the latter is advected towards smaller scales, a fraction of the energy cascades towards larger scales to maintain the balance in each triad. As discussed in the introduction, the justification for an inverse cascade of energy in three-dimensional rotating turbulence is not so straightforward since the conservation laws change in three dimensions. Nevertheless, a similar argument could be made in the case of three dimensional flows based on helicity and the possibility of an inverse cascade of energy may be alluded to, as has been explained in \cite{Briss73,Wal92}. In fact, it has been argued in \cite{Wal92} that interactions between three helical modes of the same polarization $s$, will lead to an inverse energy cascade for the same reason as has been postulated by Kraichnan \cite{Kraichnan67} and \textit{Fj{\o}rtoft} in the two dimensional case. Numerical simulations of three dimensional flows, where the non-linear interactions have been restricted to identically polarized wave numbers in all triads, further corroborate the aforementioned argument \cite{biferale_11}.

In the previous subsection we have summarized theories that clearly vindicate the notion of a gradual transfer of energy towards the slow manifold, without being able to formally account for the exact coupling between the 3D and the 2D modes. However, once the energy is in the 2D modes and if the coupling between the 2D and the 3D modes is weak, one can naively expect an inverse cascade to develop as in the case of 2D turbulence. The strength of the coupling between the 2D and the 3D modes {has been} studied by \cite{Babin96} and also by \cite{lydia08} using numerical simulations. In the case of infinitely small $\mathcal{R}o$ and in a periodic box, the 2D modes are expected to decouple from the 3D modes and evolve under their own dynamics. This is in agreement with the evidence of decoupling between the 2D and the 3D modes that was observed in numerical simulations of freely decaying rotating turbulence \cite{Teitel11} and of ideal helical rotating flows \cite{ideal_rot}. However, note that some authors claim {that} these modes never decouple in infinite domains \cite{Cambon04}.

The decoupling is further illustrated below based on the presentation in \cite{lydia08} and extended to consider the flux of energy interchanged between the 2D and the 3D modes. It is important to note that Refs.~\cite{lydia08,ideal_rot,Teitel11} studied rotating flows in the absence of forcing, thereby making a case for analyzing a completely decoupled set of equations for the 2D and the 3D modes; however, Ref.~\cite{Bourouiba2011} considers the effect of forcing.

We write wavenumbers in three-dimensional Fourier space using cylindrical coordinates, ${\bf k} = ({\bf k}_\perp, \bf k_\parallel)$, with ${\bf k}_\perp = (k_x,k_y,0) = (\rho_k, \phi_k)$, ${\bf k_{||}} = (0,0,k_z)$ and $k = |{\bf k}|$. We denote the 2D modes in Fourier space as ${\bf u}_{2D}({\bf k}_\perp)$, and the 3D or wave modes as ${\bf u}_{3D}({\bf k})$. Following \cite{lydia08}, wave vectors are decomposed as follows:
\begin{align}
 W_k &:= \{{\bf k} \text{ s.t. }|{\bf k}| \ne 0 \text{ and } k_{||}\ne 0\}, \nonumber\\
 V_k &:= \{{\bf k} \text{ s.t. } |{\bf k}| \ne 0 \text{ and } k_{||}=0\}. \nonumber
\label{modaldecomp}
\end{align}
Then the velocity field ${\bf u}=(u,v,w)$ can be decomposed as:
\begin{equation}
{\bf u}({\bf k}) = \left\{ \begin{array}{ll}
 {\bf u}_{3D}({\bf k}) & \textrm{if } {\bf k} \in W_k \\
 {\bf u}_\perp({\bf k}_\perp) + w({\bf k}_\perp)\hat{z} & \textrm{if } {\bf k} \in V_k \\
\end{array} \right.
\label{velmodaldecomp}
\end{equation}
where ${\bf u}_{2D}({\bf k}_\perp) = {\bf u}_\perp({\bf k}_\perp) + w({\bf k}_\perp)\hat{z}$. Likewise, the total energy $E = \sum_{{\bf k}}|{\bf u}({\bf k})|^2/2$ may be written in terms of the modal components as $E = E_{3D} + E_{2D} = E_{3D} + (E_{\perp} + E_{w})$, where $E_{3D}= \sum_{{\bf k} \in W_k}|{\bf u}({\bf k})|^2$/2, $E_{\perp} = \sum_{{\bf k}_{\perp}}|{\bf u}_{2D}({\bf k}_{\perp})|^2/2$ and $E_w = \sum_{{\bf k}_{\perp}}|w({\bf k}_{\perp})|^2/2$.

Multiplying the spectral form of Eq.~\eqref{NSErot1} by ${\bf u}^\star({\bf k})$ and integrating over all wavenumbers in $W_k$ and $V_k$ respectively, results in two differential equations for the total energy in the wave and the slow modes as follows:
\begin{align}
d_t E_{3D} &= \Pi_{2D \to 3D} - \Pi_{3D} + \epsilon_{3D} \label{EnFlx3D},\\
d_t E_{2D} &= -\Pi_{2D \to 3D} - \Pi_{2D} + \epsilon_{2D} \label{EnFlx2D},
\end{align}
{where the $\epsilon_{3D}$ and $\epsilon_{2D}$ terms refer to the corresponding components of the forced energy injection,} 
{and 3D (2D) refers to $k_{||}\not= 0$ ($k_{||}= 0$), as stated before.} Equations \eqref{EnFlx3D} and \eqref{EnFlx2D} are congruous to the equations derived in \cite{lydia08}. When positive, the term $\Pi_{3D}$ refers to the 3D energy that is transferred to small scales and dissipated per unit of time (thus balancing the $\epsilon_{3D}$ term) and results from triadic interactions that move energy from the 3D modes to the 3D modes (resonant interactions involving three fast modes or those between two fast modes and one slow mode). Similarly, the term $\Pi_{2D}$ results from all triadic interactions that move energy from the 2D modes to the 2D modes and, when positive, it's net effect is to balance the injection of energy per unit time in the 2D modes. Finally, $\Pi_{2D\to3D}$ is the flux of energy across $k_{||}=0$ in Fourier space, i.e., energy going from the 2D to the 3D modes when $\Pi_{2D\to3D}(t)>0$. This term is expected to be ${\mathcal O}({\mathcal R}o)$ \cite{lydia08}, and as a result, in the limit of zero $\mathcal{R}o$, the slow manifold decouples from the wave modes and the energy equations \eqref{EnFlx3D} and \eqref{EnFlx2D} are as follows:
\begin{align}
d_t E_{3D} &= -\Pi_{3D} + \epsilon_{3D}, \nonumber \\
d_t E_{2D} &= -\Pi_{2D} + \epsilon_{2D} . \nonumber 
\end{align}
It may be emphasized that, in this limit, $\Pi_{2D}$ only involves triadic interactions between slow modes and $\Pi_{3D}$ involves interactions between fast modes. Moreover, the equation for the evolution of the 2D energy further decouples into equations for $E_\perp$ and $E_w$ (see, e.g., \cite{lydia08}).

\section{Numerical simulations and subgrid scale model\label{sec:methods}}

\subsection{Large eddy simulations}

We integrate the Navier-Stokes Eq.~\eqref{NSErot1} in a rotating frame of reference using a parallel pseudo-spectral code with periodic boundary conditions \cite{hybridGHOST11}. A second order Runge-Kutta method is used to evolve the equations in time and no dealiasing is done because a LES is used; so the maximum resolved wavenumber is $k_c = N/2$, where $N$ is the linear resolution. As large scale separation between the box size and the forcing scale is essential to study inverse cascades with reasonably large values of the Reynolds number, we use large eddy simulations (LES). The sub-grid scale model is such that the wavenumbers below a cut-off wavenumber, $k_c$, are resolved explicitly whereas larger wavenumbers are modeled based on energy and helicity contributions to the eddy viscosity and the eddy noise terms in the EDQNM equations. For completeness, the model is summarized below.

First, the larger resolved scales are computed by integrating the following equation:
\begin{align}
&& \left[ \frac{\partial}{\partial t} + k^2 \left(\frac{1}{{\mathcal R}e} + \nu_{k|k_c}\right) \right] 
    {\bf u}_{\alpha}({\bf k},t) & = T^{<}_{\alpha}({\bf k},t) \nonumber \\ 
{} && - \frac{1}{{\mathcal R}o} P_{\alpha\beta} \, \varepsilon_{\beta\gamma\zeta}{\bf u}_{\zeta}({\bf k},t)
    + {\bf f}_{\alpha}({\bf k},t), \label{modelRES}
\end{align}
which is basically the Fourier transform of Eq.~\eqref{NSErot1} (barring the newly introduced subgrid model term, $\nu_{k|k_c}$). Here the greek subindices denote cartesian components of the vectors and tensors and Einstein summation convention is assumed. The term $T^{<}_{\alpha}({\bf k},t)$ is the Fourier transform of the nonlinear term in Eq.~\eqref{NSErot1} over all modes with $k<k_c$. In other words, it represents the nonlinear transfer that arises from the convolution sum over a truncated triadic domain, ${\bf k} + {\bf p} + {\bf q} = 0$, $k,p,q < k_c$. This term is computed using the pseudo-spectral method. The eddy viscosity term, $\nu_{k|k_c}$ is associated with the subgrid model and is computed based on parameters that are modeled from the unresolved scales. $P_{\alpha\beta}({\bf k}) =\delta_{\alpha\beta} - \frac{k_{\alpha}k_{\beta}}{k^2}$ is the projector operator on the solenoidal velocity field and $\mathbf{\varepsilon}$ is the antisymmetric tensor associated with the curl operator (Levi-Civita symbol).

The isotropic energy spectrum $E(k,t)$ and the helicity spectrum $H(k,t)$ up to wavenumber $3k_c$ (including unresolved scales) are then obtained through data fitting and extrapolation from the resolved scales. Next, the isotropic energy spectrum $E(k,t)$ for the unresolved scales is evolved based on the following:
\begin{align}
(\partial_t + 2\nu k^2)E(k,t) &= -2\nu_{k|k_c}k^2E(k,t) - 2\tilde{\nu}_{k|k_c}k^2H(k,t) \nonumber \\ 
    &+ T^{<}_{E}(k,t) + \frac{\hat{T}^{pq}_{E}(k,t)}{4\pi k^2} . \label{modelLIN}
\end{align}
An equivalent balance equation for the unresolved helicity spectrum $H(k,t)$ is solved if the helicity of the flow is non-zero (note: when $H\equiv 0$ , we have $\tilde \nu \equiv 0$).
Here, $\nu_{k|k_c}$ and $\tilde{\nu}_{k|k_c}$ are terms prescribed by the model as before, $T^{<}_{E}(k,t)$ represents the energy transferred to unresolved scales from the resolved scales and $\hat{T}^{pq}_{E}$ represents the energy and helicity interactions at wavenumbers $p,q > k_c$. The analytical forms of the above terms come from a two-point analysis of an integro-differential equations originating from the EDQNM closure for isotropic Navier-Stokes turbulence. Thus, our model assumes that isotropy is recovered at sufficiently small scales (smaller than the Zeman scale) as was recently shown in a large DNS of rotating turbulence \cite{JFM}. It may be noted here that the LES was able to reproduce the results of this DNS on a grid of $3072^3$ points in which the Zeman scale was resolved \cite{trieste_12}. Finally, eddy-noise (upscaling of energy towards the resolved scales from the unresolved scales) is added to the velocity field based on a reconstruction of Eq.~\eqref{modelLIN}. The reader is referred to {\cite{Baerenzung08} (see  Eqs.~(27) and (28))} for further explanation.

\subsection{Description of the runs}

\begin{table}
\caption{
Table of the runs with the total relative helicity of the flow $\rho_H$, the anisotropy exponent $\beta$, 
 the forcing scale Rossby and Reynolds numbers, $Ro_f$ and $Re_f$, the energy injection rate $\epsilon$, and the power law index in the inverse cascade range of the spectrum of the 2D modes. TG, ABC, RND, and ANI respectively stand for Taylor-Green, ABC, random, and random anisotropic forcing. {Note that $\rho_H$ is the relative helicity of the flow at the time when the inverse cascade starts, i.e. at t=0 in the run with rotation. {All runs use a grid with $N=256$ points, a forcing wavenumber $k_f=40$, an imposed rotation $\Omega=35$, and a kinematic viscosity $\nu=2\times 10^{-4}$.}
}}
\begin{ruledtabular}  \begin{tabular}{ccccccc}
Run   &  $\rho_H$& $\beta$&  ${\mathcal R}o_f$&${\mathcal R}e_f$ &$\epsilon$ & index         \\
\hline
TG    & $8 \times 10^{-3}$    &   --   &   {0.045}                & {390}                 & {0.030}          & $\approx-3$   \\
RND1  & $9 \times 10^{-3}$    &   --   &  {0.045}                & {390}                &  {0.047}         & $\approx-3$   \\
RND2  & $8 \times 10^{-2}$    &   --   &  {0.044}                & {390}                 &    {0.050}       & $\approx-3$   \\
RND3  & $5 \times 10^{-1}$   &   --   &   {0.046}                & {420}                 &   {0.047}        & $\approx-3$   \\
RND4  & $7 \times 10^{-1}$ & -- &   {0.044}                & {420}                 &  {0.047}         & $\approx-3$   \\
ANI1  & $1 \times 10^{-2}$    &   1    &   {0.045}                & {400}                 &  {0.010}         & $\approx-3$   \\
ANI2  & $8 \times 10^{-3}$    &   2    &  {0.045}                & {400}                 &  {0.010}         & $\approx-3$   \\
ANI3  & $8 \times 10^{-3}$   &   3    &   {0.045}                & {420}                 &   {0.007}        & $\approx-5/3$ \\
ANI4  & $7 \times 10^{-1}$ & 3  &   {0.045}                & {420}                 &   {0.006}        & $\approx-5/3$\\
ABC   & $7 \times 10^{-1}$ & -- &   {0.050}                &  {470}                &  {0.090}         & $\approx-5/3$ \\
\end{tabular} \end{ruledtabular}
\label{tab1} \end{table}

Since the main aim of this paper is to study the inverse cascade of energy, all simulations are forced at high wavenumbers $k_f$ to ensure sufficient scale separation between the forcing and the box scale. We explore different forcing functions in order to consider the effects of spectral anisotropy, number of components in the forcing, and the role of helicity on the dynamics of the flow at large scales (see table \ref{tab1}).

For the Taylor-Green (TG) run, ${\bf f}$ in Eq.~\eqref{NSErot1} is the following \cite{Taylor37}:
{\setlength\arraycolsep{2pt}
\begin{eqnarray}
{\bf f}_{\rm TG} &=& f_0 \left[ \sin(k_{TG} x) \cos(k_{TG} y) \cos(k_{TG} z) \hat{x}  \right. {} \nonumber \\
&& {} \left. - \cos(k_{TG} x) \sin(k_{TG} y) \cos(k_{TG} z) \hat{y} \right], 
\label{eq:TG}
\end{eqnarray}}where $f_0 = 5.0$ is the forcing amplitude. Such a forcing function injects zero net helicity in the flow, excites only two components of the flow ($u$ and $v$, although $w$ also grows with time as a result of pressure forces). ${\bf f}_{TG}$ injects energy only into a few 3D modes (no energy is injected directly into modes with $k_\parallel = 0$). The expression in Eq.~\eqref{eq:TG} has many symmetries that are preserved during the evolution of the Navier-Stokes equation. To break these symmetries and reach a turbulent steady state faster, a superposition of two TG forcing functions acting at {$k_{TG}= 21$ and $22$} was used. Since TG forcing involves products of three modes in Fourier space, the effective forcing wavenumber is $k_f = \sqrt{3} \min\{{k_{TG}}\} \approx 40$, while the projection of the forcing wavenumber into the plane of 2D modes is $k_{\perp,f} = \sqrt{2} \min\{{k_{TG}}\} \approx 30$.

In order to study the effect of helicity, we also performed simulations using the Arn'old-Childress-Beltrami (ABC) forcing \cite{Arnold72}:
{\setlength\arraycolsep{2pt}
\begin{eqnarray}
{\bf f}_{\rm ABC} &=& f_0 \left\{ \left[B \cos(k_f y) +  C \sin(k_f z) \right] \hat{x}  \right. \nonumber \\
&& {} + \left[A \sin(k_f x) + C \cos(k_f z) \right] \hat{y} \nonumber \\
&& {} + \left. \left[A \cos(k_f x) + B \sin(k_f y) \right] \hat{z} \right\},
\label{eq:ABC}
\end{eqnarray}}with $A=0.9$, $B=1$ and $C=1.1$. ABC forcing is an eigenfunction of the curl operator and injects maximum helicity {(i.e., ${\bf f}_{ABC}$ and $\nabla \times {\bf f}_{ABC}$ are co-linear).}
 When the flow is forced using this type of forcing function, turbulence develops only after an instability sets in \cite{Podvigina94}. To speed up the onset of turbulence, we forced the flow with a superposition of two ABC flows at $k_f = 40$ and $41$. Henceforth, $k_f$ refers to the minimum of the two forcing wavenumbers. Note that the ABC forcing excites only two 2D modes in the Fourier shell with $k=k_f$ (in the $k_x$ and $k_y$ axis in Fourier space) and one 3D mode (in the $k_\parallel$ axis).

We also used two types of randomly generated isotropic forcings. In the first type (labeled RND in table \ref{tab1}), all modes in spherical Fourier shells between {$k_f=40$} and $41$ were fired with the same amplitude but random phases. The method described in \cite{Pouquet78} was used to correlate phases and change the helicity of the forcing from zero to maximal. This results in isotropic forcing independent of the amount of helicity. As a result, more energy is injected into 3D modes compared to 2D modes. The second random forcing (labeled ANI in table \ref{tab1}) corresponds to a case in which we {introduce a new parameter, $\beta$ in order to} control the extent of anisotropy in the forcing. We define a new forcing function by multiply{ing} each mode in the RND forcing by a factor that concentrates the effective forcing near the slow manifold,
\begin{equation}
{\bf f}_{\rm ANI}({\bf k}) = \left(1-\frac{k_z}{k_f}\right)^\beta {\bf f}_{\rm RND}({\bf k}).
\label{FANI} \end{equation}
Note that $\beta=0$ corresponds to isotropic forcing. Refer to table \ref{tab1} for the different values of $\beta$ used in the simulations. 

The simulations were started from a flow at rest and without rotation and integrated up to ten large-scale turnover times ($\tau_f=L_f/U \approx 10$) until a turbulent steady state with a well developed direct energy cascade was attained. Next, at a time arbitrarily re-labeled $t=0$, rotation was turned on and the simulation was continued for at least {250} $\tau_f$ turnover times. A fully developed inverse cascade of energy was observed by this time in all the runs mentioned in table \ref{tab1}.

\subsection{Anisotropic spectra}

To study power laws in the 
{resulting}
inverse cascade range of the simulations we refer to both isotropic and anisotropic energy and helicity spectra. The decomposition of the total energy into energy in 2D and 3D modes, $E_{2D}$ and $E_{3D}$ described above can be extended to spectral densities as follows, based on the definitions presented in \cite{JFM}.

The isotropic total energy spectrum is computed in the simulations as:
\begin{equation}
E(k) = \frac{1}{2} \sum_{k\le |{\bf k}|< k+1} |{\bf u}({\bf k})|^2, 
\end{equation}
and is such that the total energy is $E=\sum_k E(k)$. We can also define an axisymmetric energy spectrum:
\begin{equation}
e(k_\perp,k_\parallel)= \frac{1}{2}
    \sum_{\substack{
          k_{\perp}\le |{\bf k}\times \hat {\bf z}| < k_{\perp}+1 \\
          k_{\parallel}\le k_z < k_{\parallel}+1}} |{\bf u}({\bf k})|^2
    = e(k, \theta_k), 
\label{etheta} \end{equation}
where 
{in the latter expression,}
$\theta_k$ is the colatitude in Fourier space with respect to the rotation axis. The axisymmetric energy spectrum is such that the total energy in 2D modes is $E_{2D} = \sum_{k_\perp} e(k_\perp,k_\parallel=0) = \sum_k e(k,\theta_k=\pi/2)$. As a result, we refer to $e(k_\perp,k_\parallel=0)$ as the energy spectrum of the 2D modes.

\textit{Reduced} perpendicular and parallel spectra can then be defined as:
\begin{equation}
E(k_\perp) = \sum_{k_\parallel} e(k_\perp,k_\parallel) ,
\end{equation}
and
\begin{equation}
E(k_\parallel) = \sum_{k_\perp} e(k_\perp,k_\parallel)
\end{equation}respectively.
As for the energy, $E=\sum_{k_\perp} E(k_\perp) = \sum_{k_\parallel} E(k_\parallel)$. We then introduce the isotropic and perpendicular energy spectra of the 3D modes:
\begin{equation}
E_{3D}(k) = E(k)-e(k,\theta_k=\pi/2) ,
\end{equation}
and
\begin{equation}
E_{3D}(k_\perp) = E(k_\perp)-e(k_\perp,k_\parallel=0) .
\end{equation}

Finally, we associate energy fluxes with the energy spectra $E(k)$, $E(k_\perp)$, and $E(k_\parallel)$. These are defined from the transfer functions as follows:
\begin{equation}
T(k) = - \sum_{k\le |{\bf k}|< k+1} {\bf u}^\star({\bf k}) \cdot \widehat{\left({\bf u} \cdot \nabla {\bf u}\right)}_{\bf k} ,
\end{equation}
\begin{equation}
T(k_\perp) = - \sum_{k_\perp \le |{\bf k}\times \hat{z}|< k_\perp+1} {\bf u}^\star({\bf k}) \cdot \widehat{\left({\bf u} \cdot \nabla {\bf u}\right)}_{\bf k} ,
\end{equation}
and
\begin{equation}
T(k_\parallel) = - \sum_{k_\parallel \le k_z < k_\parallel + 1} {\bf u}^\star({\bf k}) \cdot \widehat{\left({\bf u} \cdot \nabla {\bf u}\right)}_{\bf k} ,
\end{equation}
where the superscript ~$\widehat{}$ ~denotes Fourier transformed quantities.
Then, the fluxes are as follows:
\begin{equation}
\Pi(k) = -\sum_{k'=0}^k T(k') 
\end{equation}
\begin{equation}
\Pi(k_\perp) = -\sum_{k_\perp'=0}^{k_{\perp}} T(k'_\perp), \,\,\, 
\Pi(k_\parallel) = -\sum_{k'_\parallel=0}^{k_\parallel} T(k'_\parallel)
\end{equation}
These fluxes represent energy per unit of time across spheres in Fourier space with radius $k$, cylinders with radius $k_\perp$ and planes with $k_\parallel=$ constant, respectively. In particular, note that $\Pi(k_\parallel=0)$ represents energy transferred from 2D to 3D modes when positive, and from 3D to 2D modes when negative; actually, $\Pi_{2D \to 3D}$ is the flux across slow and fast modes defined in Eqs.~\eqref{EnFlx3D} and \eqref{EnFlx2D} in the previous section.

{Similar definitions can be written for the helicity.}

\section{Numerical results\label{sec:results}}

\subsection{Time evolution and spectra}

\begin{figure}
\begin{center}
\includegraphics[trim=0cm 6cm 0cm 7cm,clip=true,height=5.5cm,width=9.5cm]{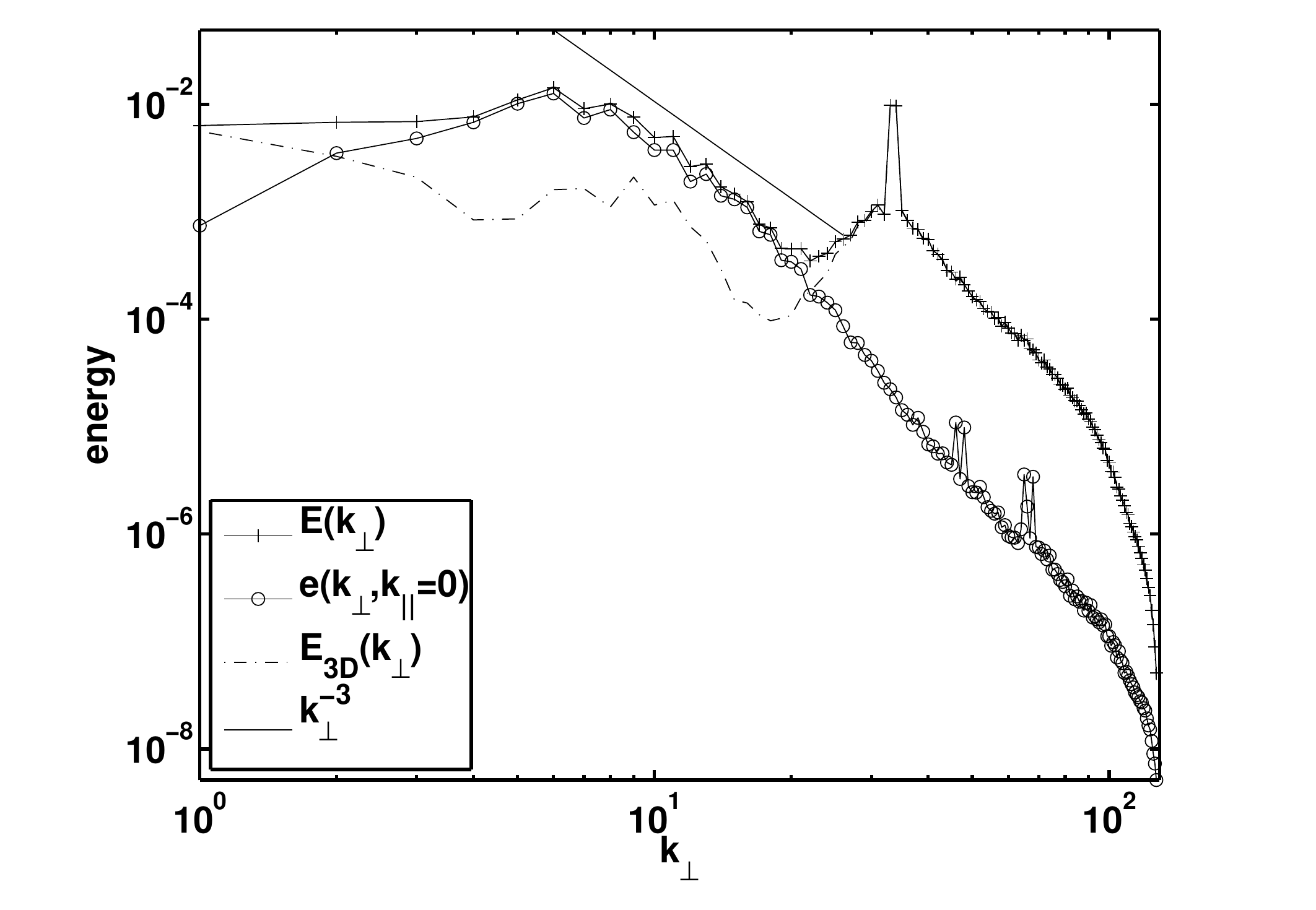}
\includegraphics[trim=0cm 6cm 0cm 7cm,clip=true,height=5.5cm,width=9.5cm]{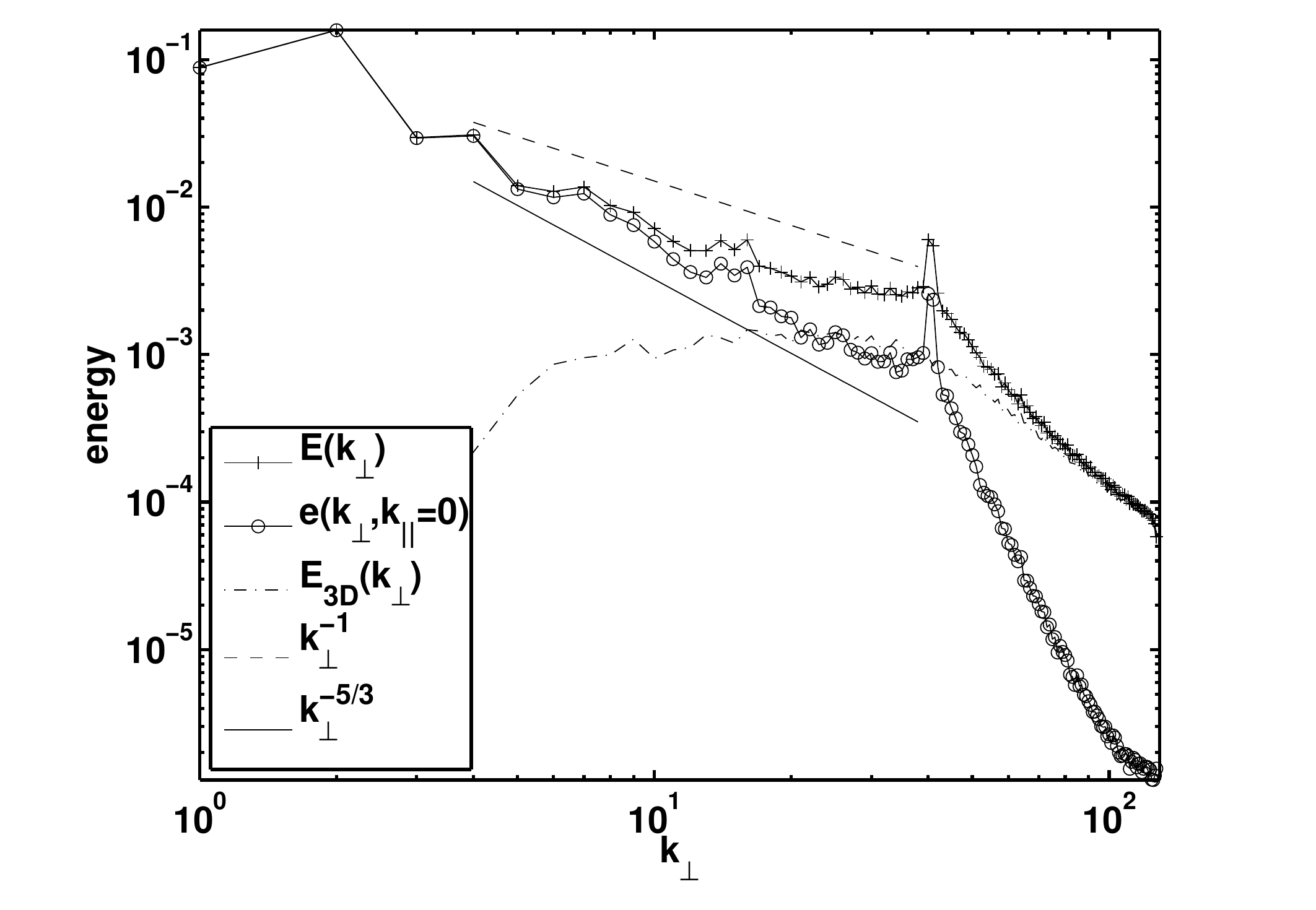}
\includegraphics[trim=0cm 6cm 0cm 7cm,clip=true,height=5.5cm,width=9.5cm]{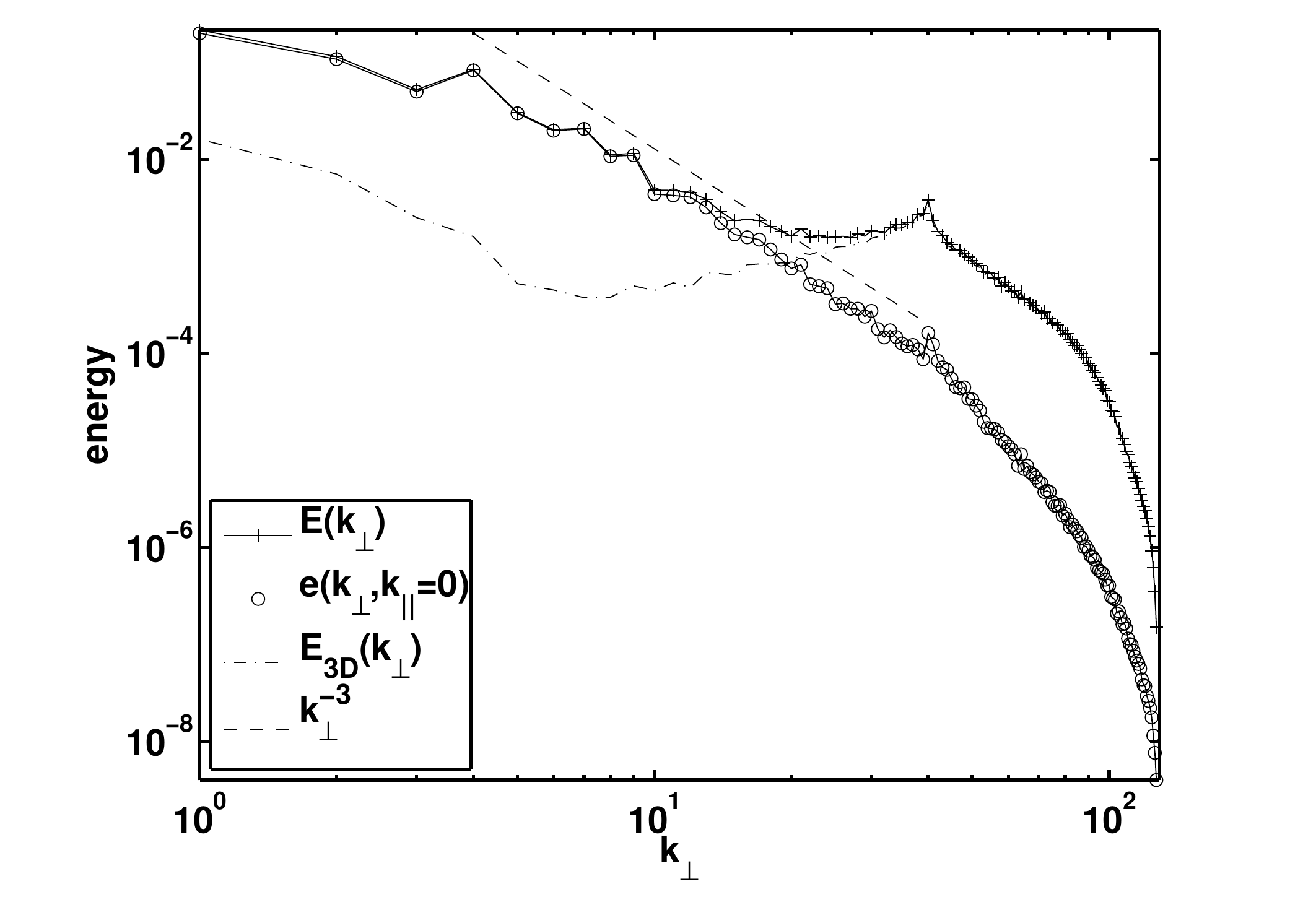}
\includegraphics[trim=0cm 6cm 0cm 7cm,clip=true,height=5.5cm,width=9.5cm]{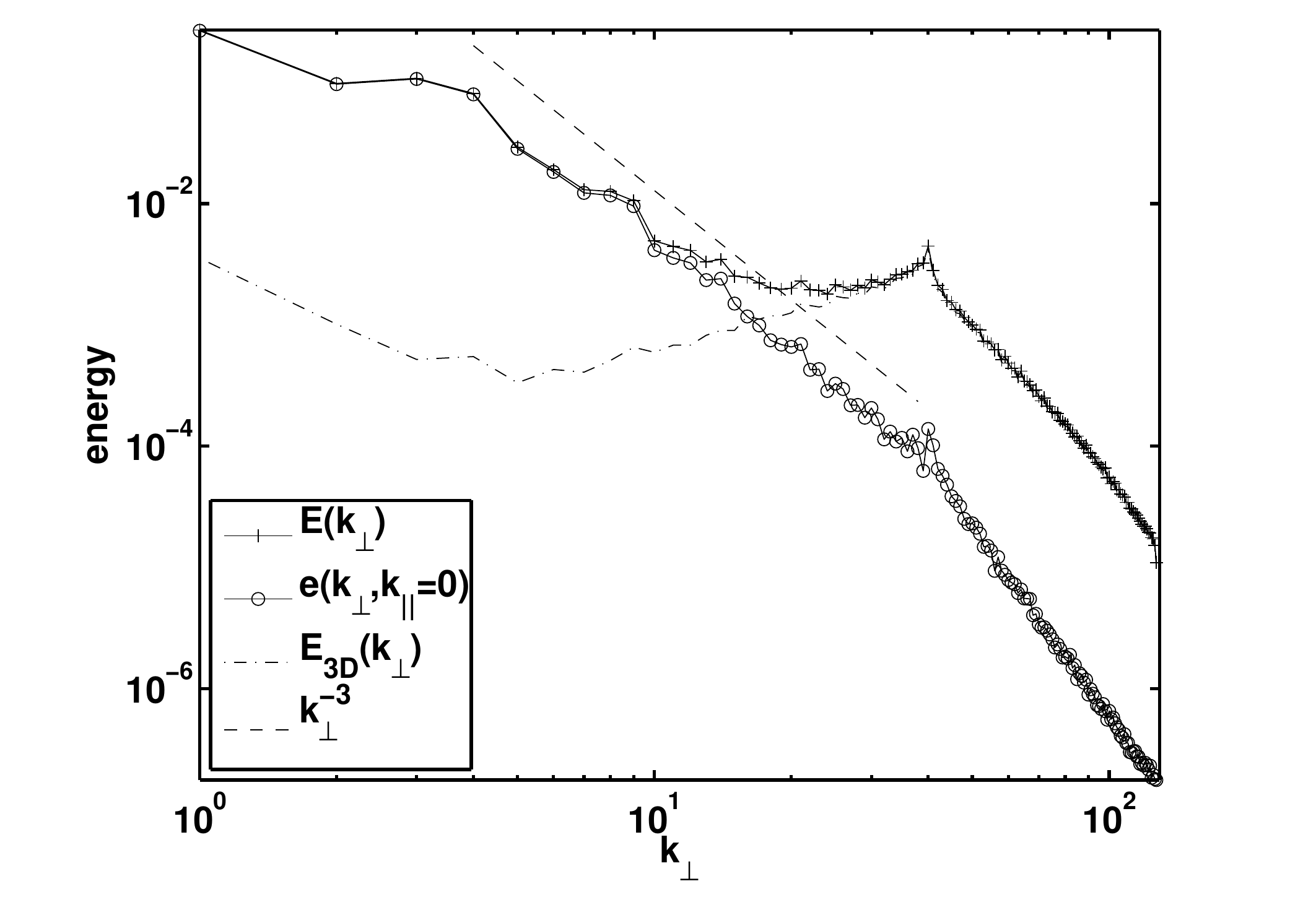}
\end{center}
\caption{$E(k_\perp)$, $E_{3D}(k_\perp)$ and $e(k_\perp,k_\parallel=0)$ at late times for TG, ABC, RND1 and RND4 forcing (from top to bottom).}
\label{fig:latetime1}
\end{figure}

At the onset, we discuss the results for simulations with TG, ABC and RND forcing with and without helicity.
The perpendicular spectrum $E(k_\perp)$, the spectrum of 3D modes $E_{3D}(k_\perp)$ and the spectrum of 2D modes $e(k_\perp,k_\parallel=0)$ at late times {are all} shown in Fig.~\ref{fig:latetime1} for the TG, ABC, RND1 and RND4 runs. The ABC run shows a spectrum $e(k_\perp,k_\parallel=0) \sim k_\perp^{-5/3}$ and $E(k_{\perp})$ $\sim k^{-1}$ at large scales. All the other runs have $e(k_\perp,k_\parallel=0) \sim k_\perp^{-3}$.
\begin{figure}
\begin{center}
\includegraphics[trim=0cm 6cm 0cm 7cm,clip=true,height=5.5cm,width=9.5cm]{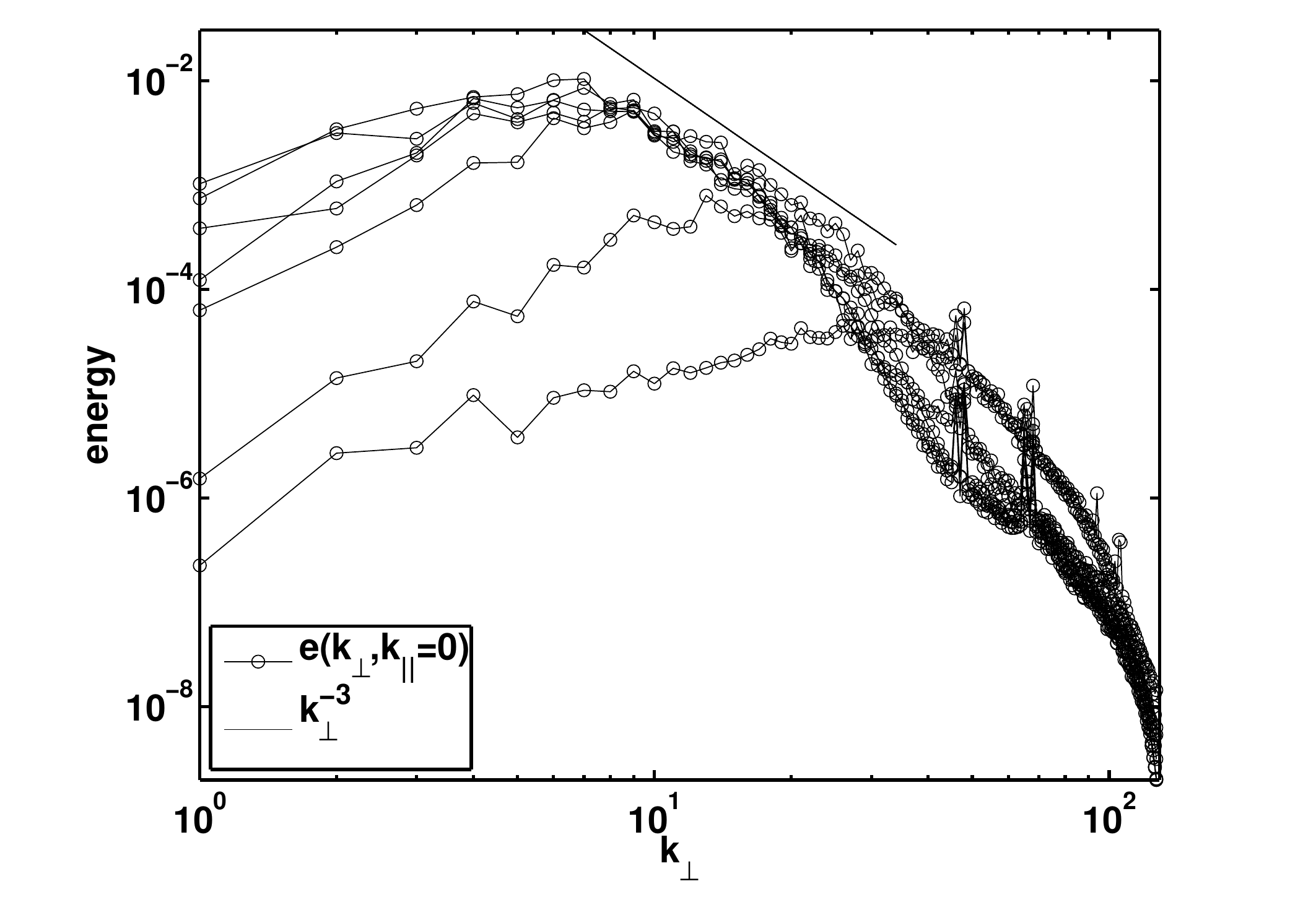}
\includegraphics[trim=0cm 6cm 0cm 7cm,clip=true,height=5.5cm,width=9.5cm]{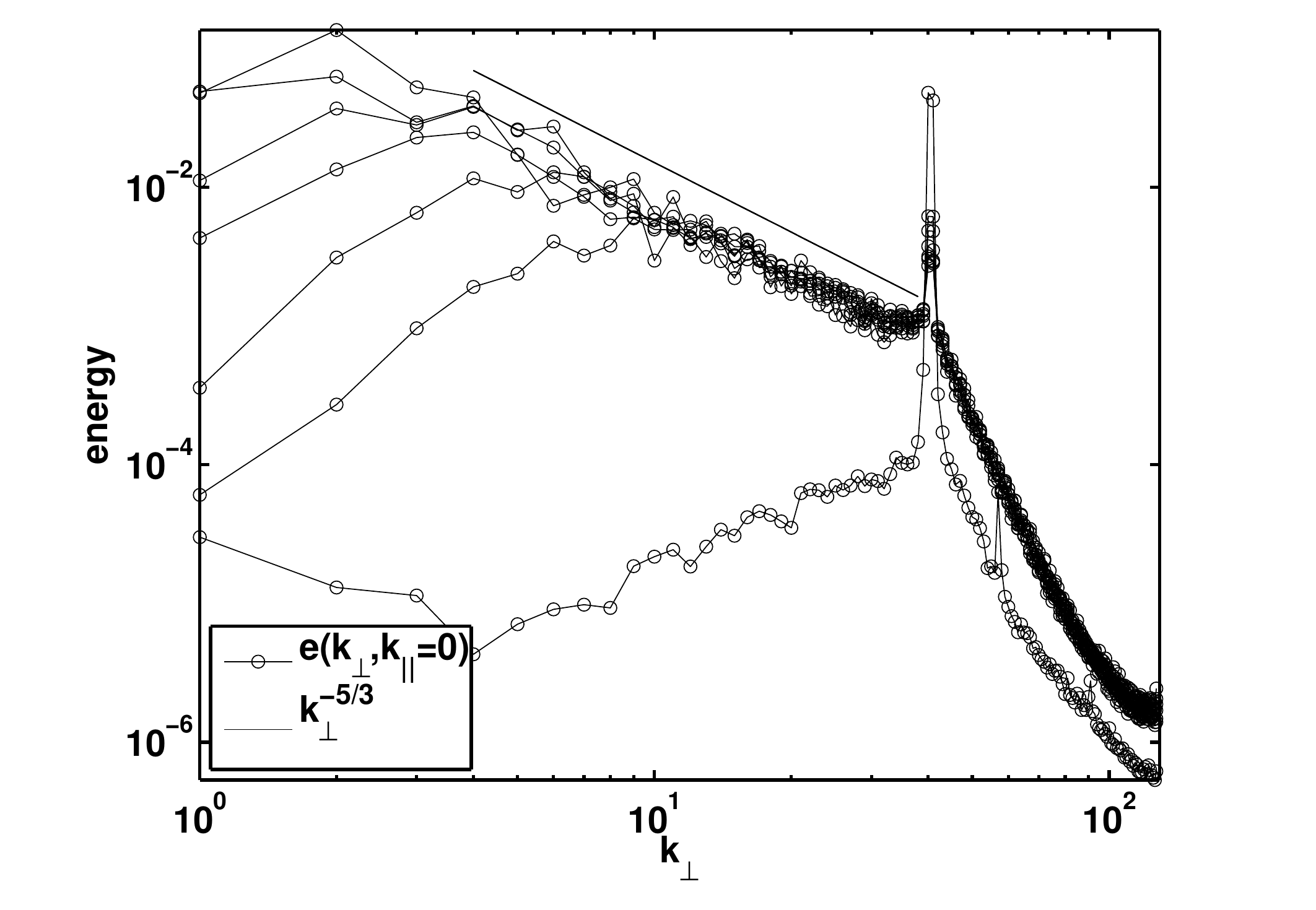}
\includegraphics[trim=0cm 6cm 0cm 7cm,clip=true,height=5.5cm,width=9.5cm]{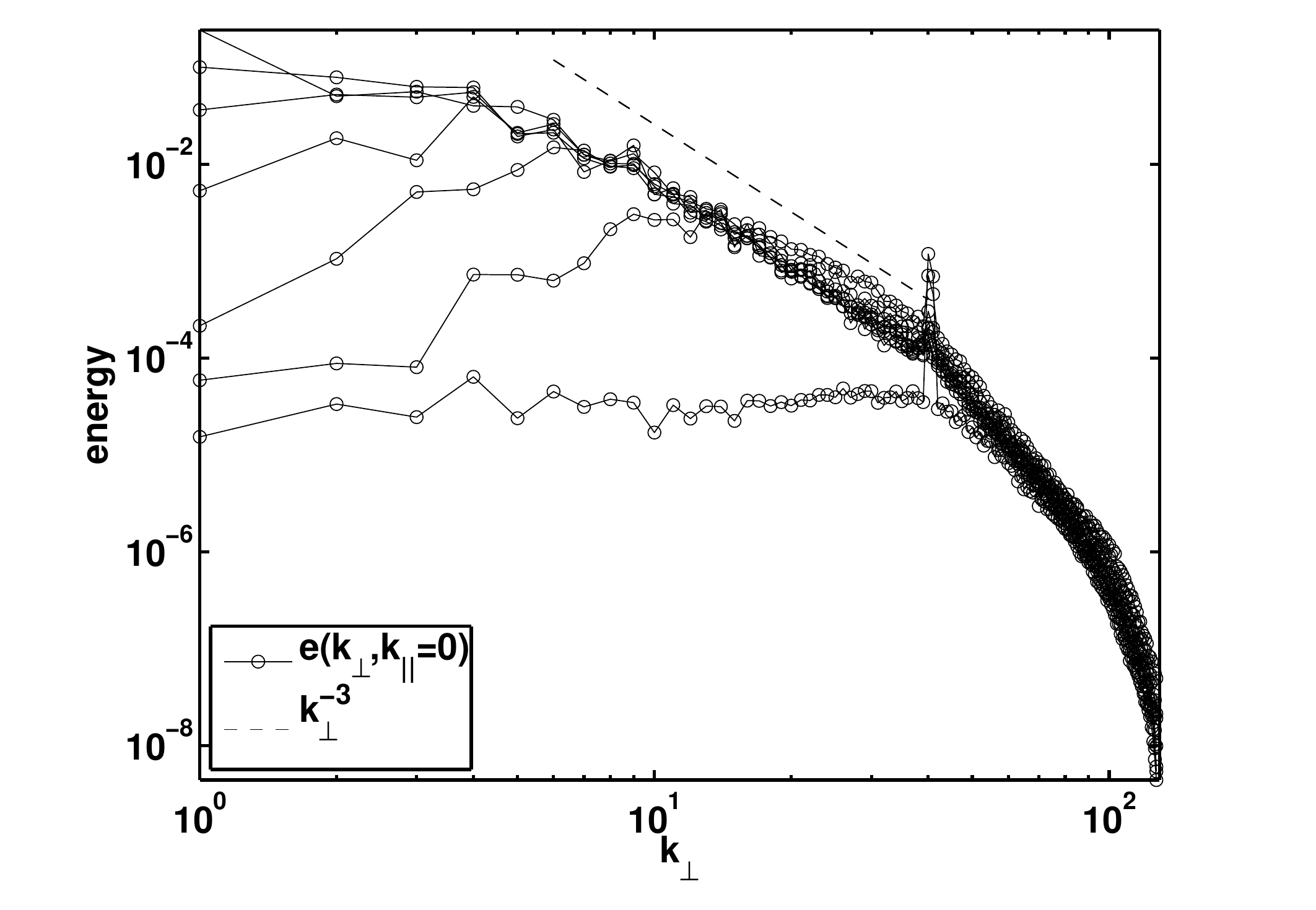}
\includegraphics[trim=0cm 6cm 0cm 7cm,clip=true,height=5.5cm,width=9.5cm]{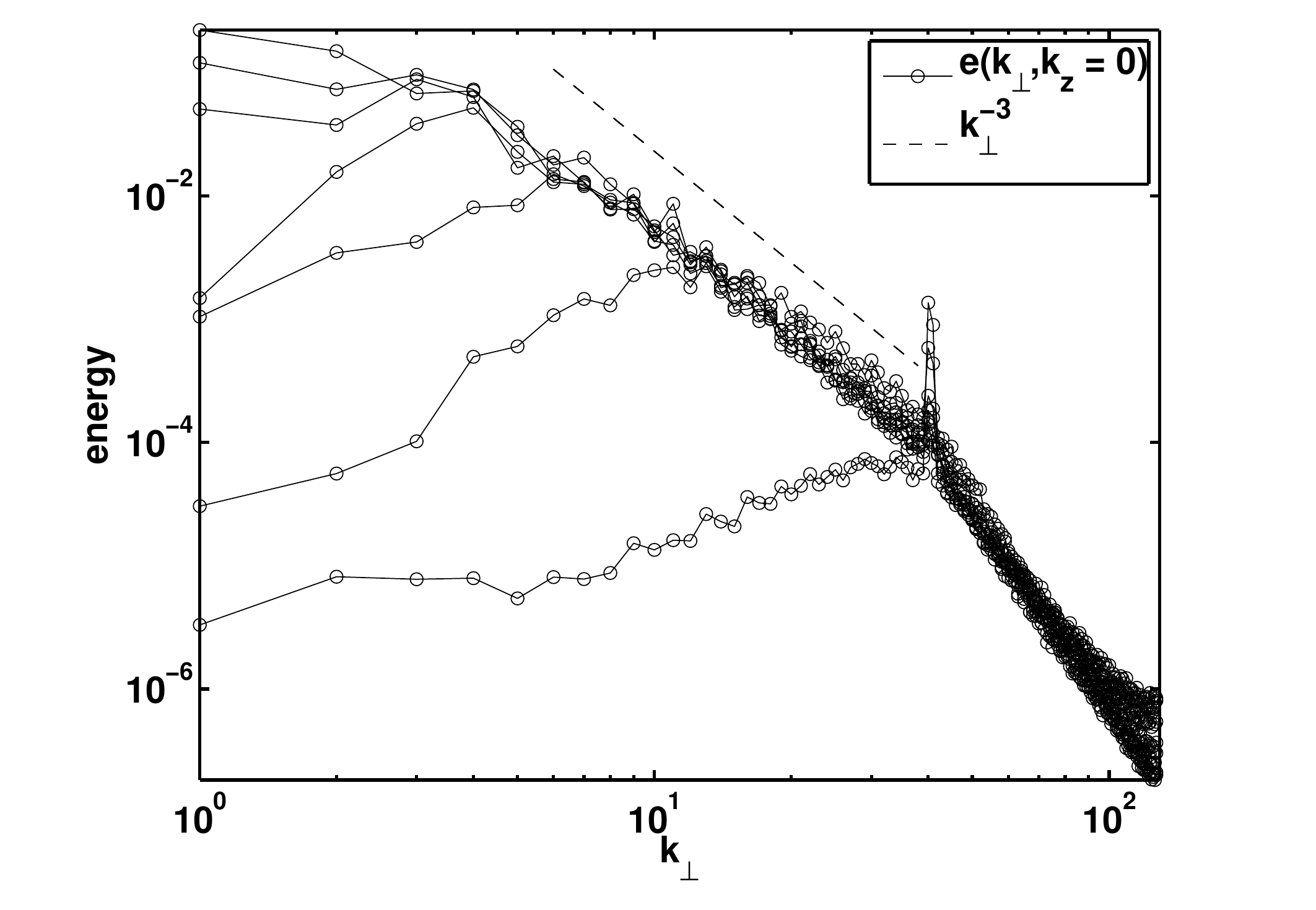}
\end{center}
\caption{Time evolution of the spectrum of the 2D energy $e(k_\perp,k_\parallel=0)$ for run TG, ABC, RND1 and RND4 (top to bottom) upto 250 turn-over times at intervals of 35.7 turn-over times. }
\label{fig:specevol1}
\end{figure}

The time evolution of the spectrum of the 2D energy $e(k_\perp,k_\parallel=0)$ for the same runs as in Fig.~\ref{fig:latetime1} is shown in Fig.~\ref{fig:specevol1}, from $t=0$ to $250\tau_f$ turn-over times at intervals of roughly $35.7\tau_f$ turn-over times.

\begin{figure}
\begin{center}
\includegraphics[trim=0cm 6cm 0cm 7cm,clip=true,height=5.5cm,width=9.5cm]{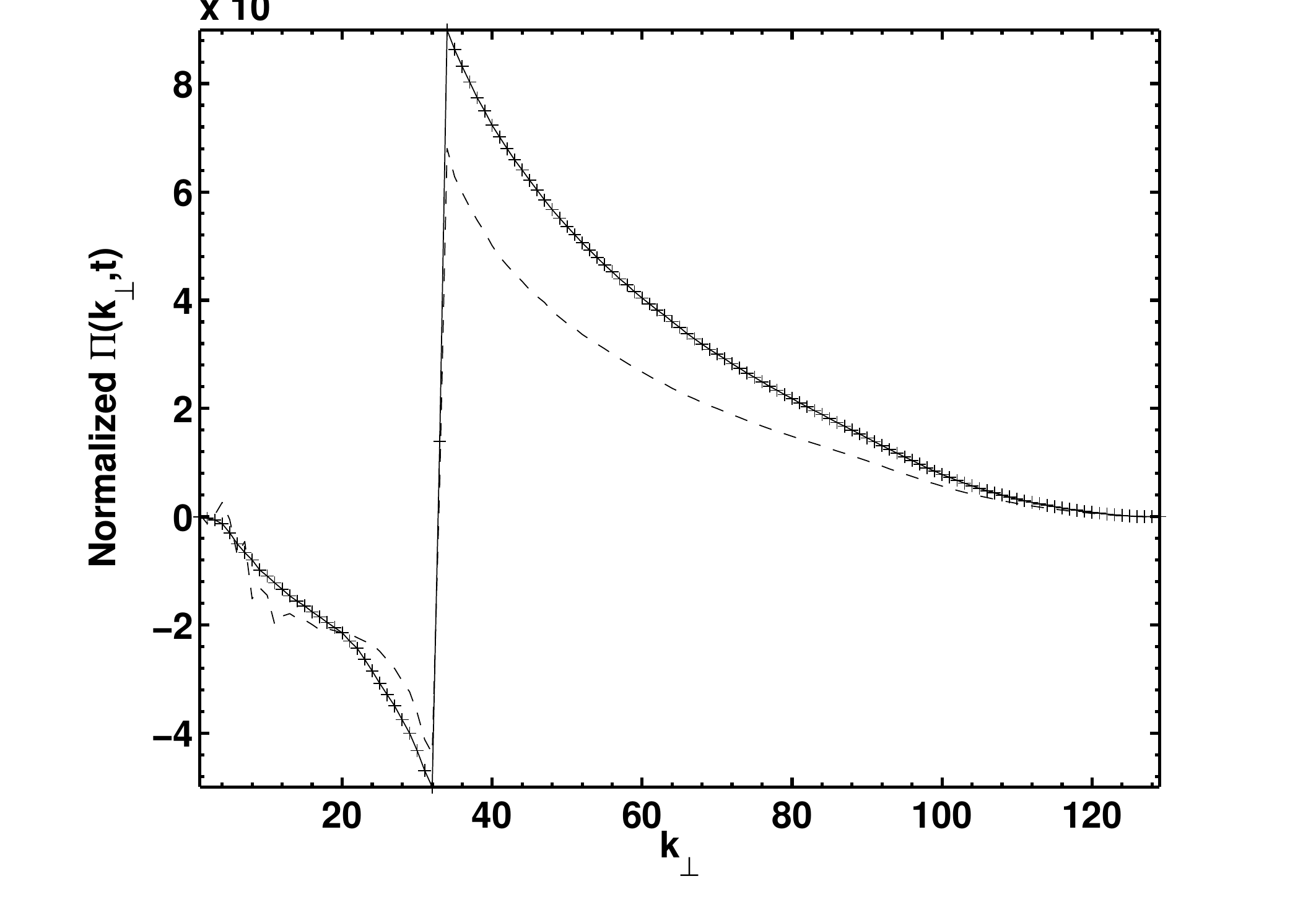}
\includegraphics[trim=0cm 6cm 0cm 7cm,clip=true,height=5.5cm,width=9.5cm]{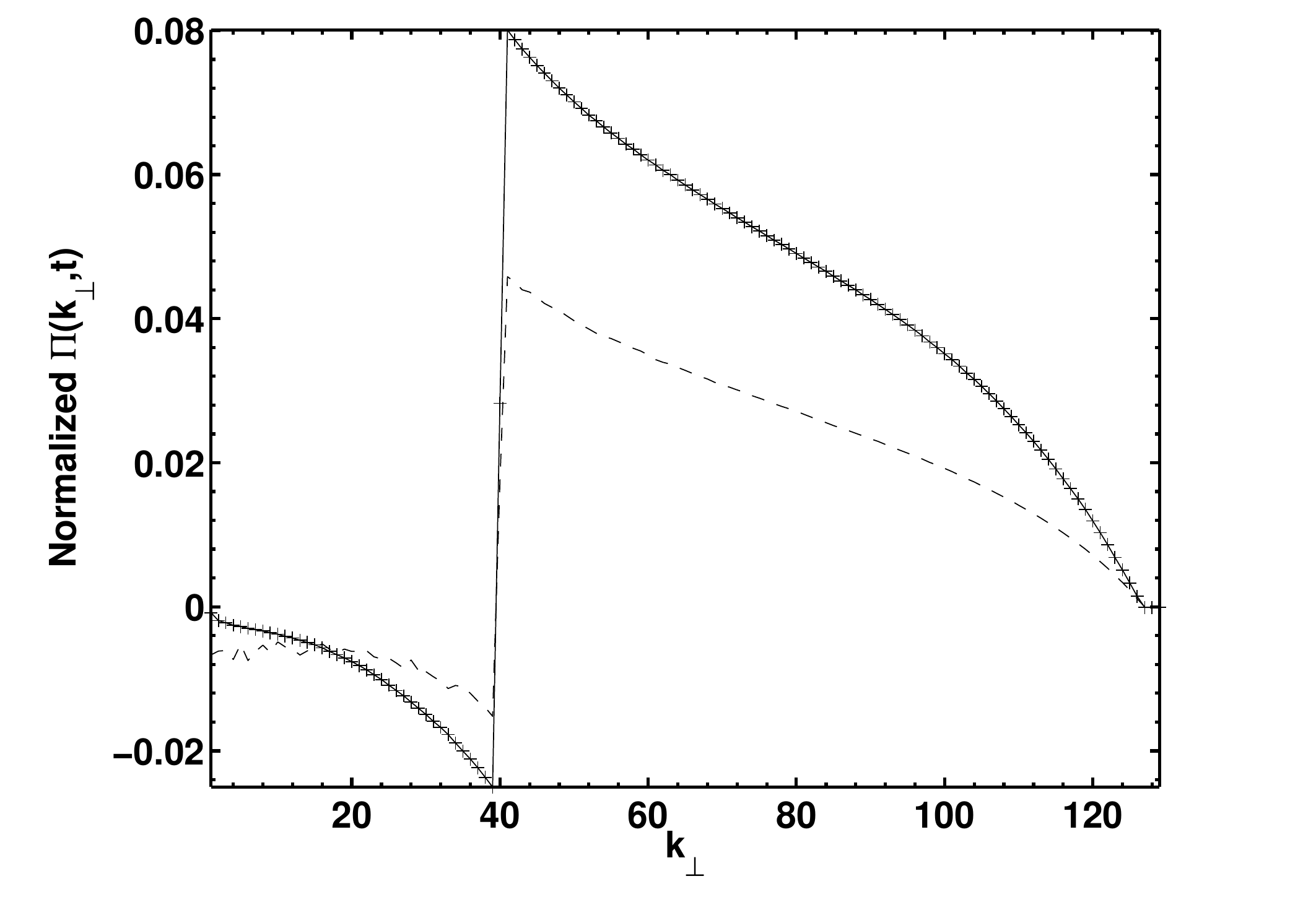}
\includegraphics[trim=0cm 6cm 0cm 7cm,clip=true,height=5.5cm,width=9.5cm]{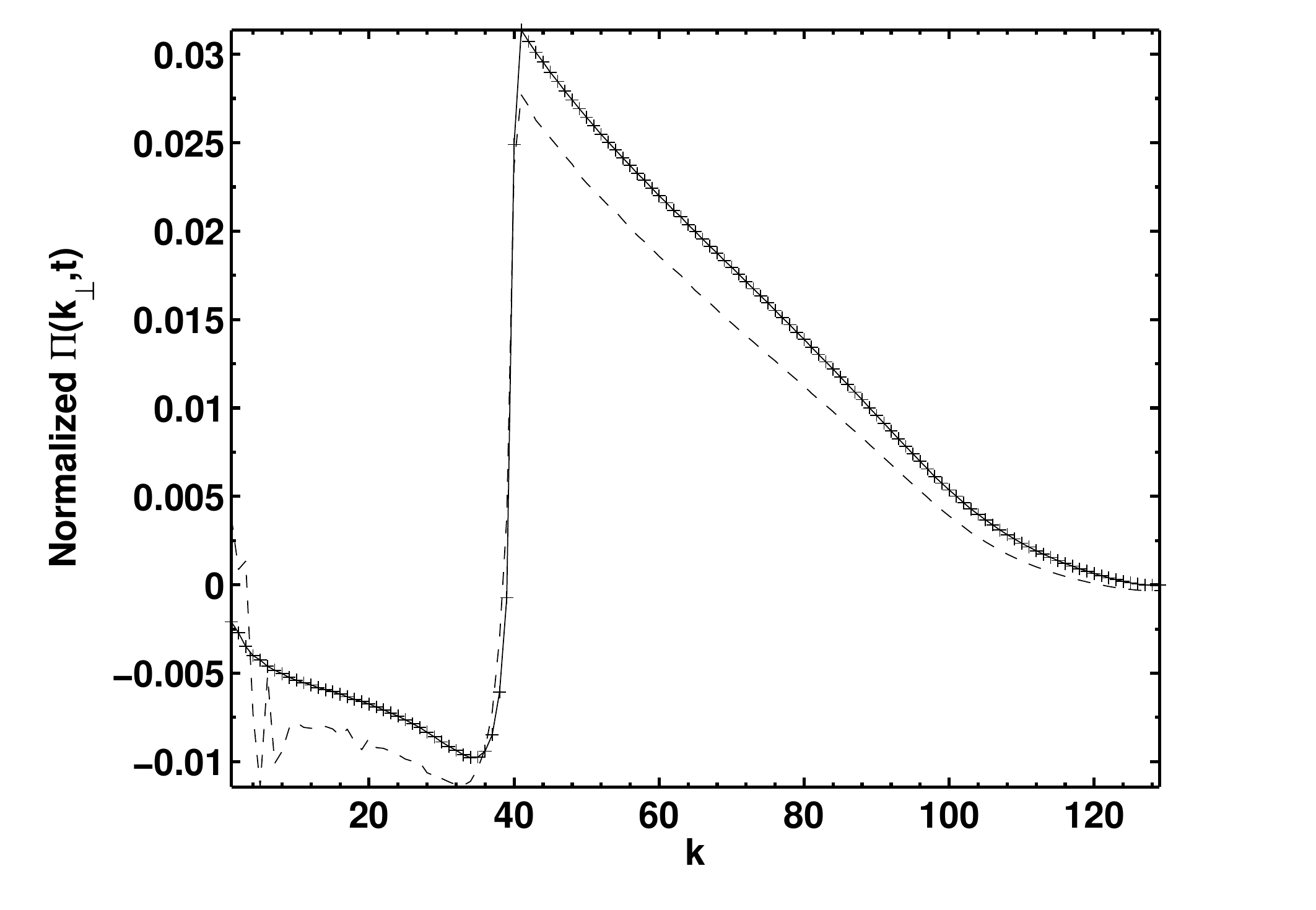}
\end{center}
\caption{Energy flux $\Pi(k_\perp)$ for runs TG, ABC and RND1 (from top to bottom). Solid lines are time averaged while dashed are instantaneous fluxes at late times. The fluxes are normalized to the value at the forcing wavenumber.}
\label{fig:flux1}
\end{figure}

The build up of energy at large scales observed in the spectra is associated with an inverse cascade of 2D energy in the presence of rotation. This can be verified from the energy flux that shows a positive range at wavenumbers larger than $k_f$ (associated with a direct cascade of energy) and a negative range at wavenumbers smaller than $k_f$ (associated with the inverse cascade). Figure \ref{fig:flux1} shows $\Pi(k_\perp)$ for runs TG, ABC, and RND{1}. The same behavior is observed in the isotropic flux $\Pi(k)$ (not shown).

\begin{figure}
\begin{center}
\includegraphics[trim=0cm 6cm 0cm 7cm,clip=true,height=5.5cm,width=9.5cm]{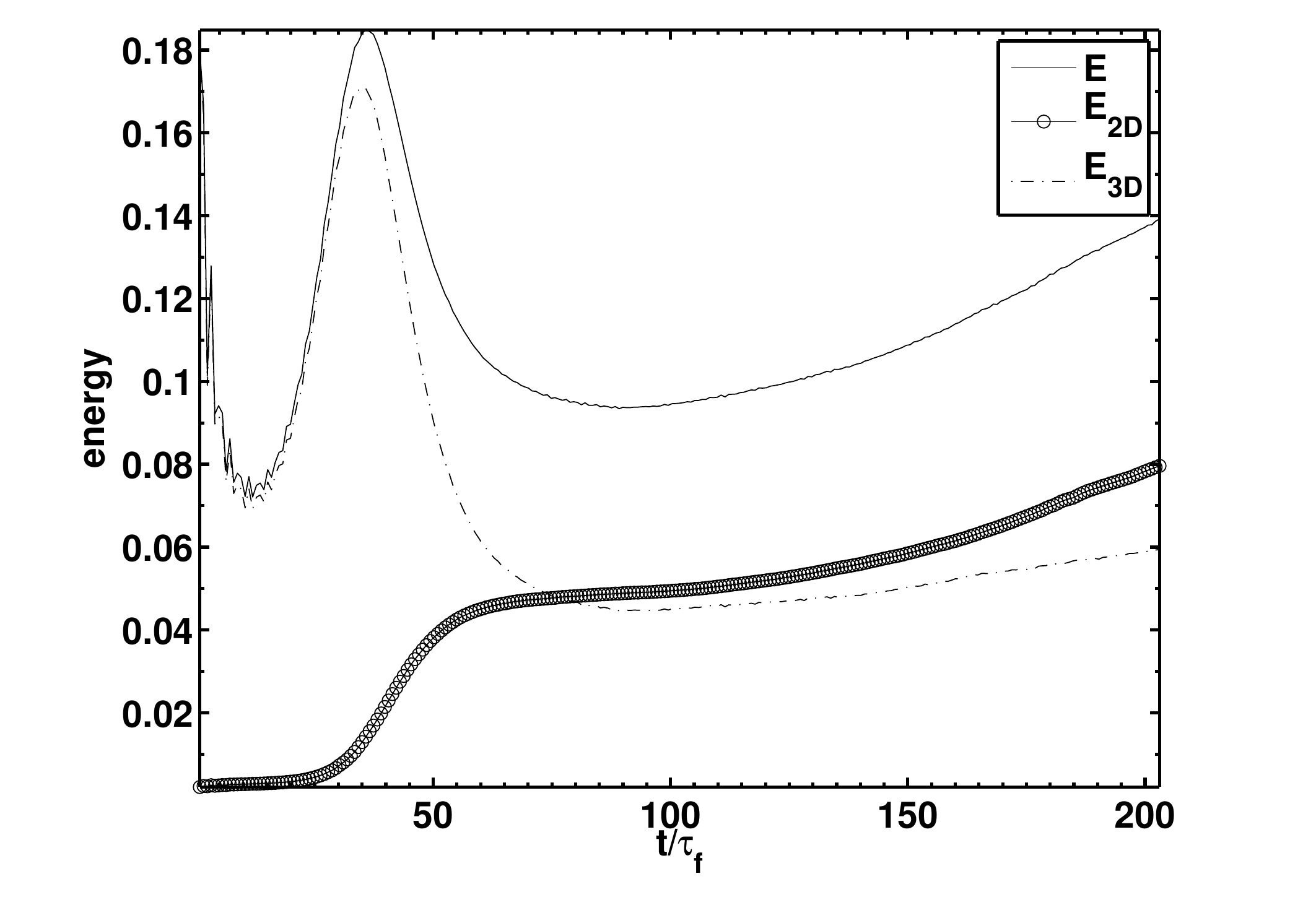}
\includegraphics[trim=0cm 6cm 0cm 7cm,clip=true,height=5.5cm,width=9.5cm]{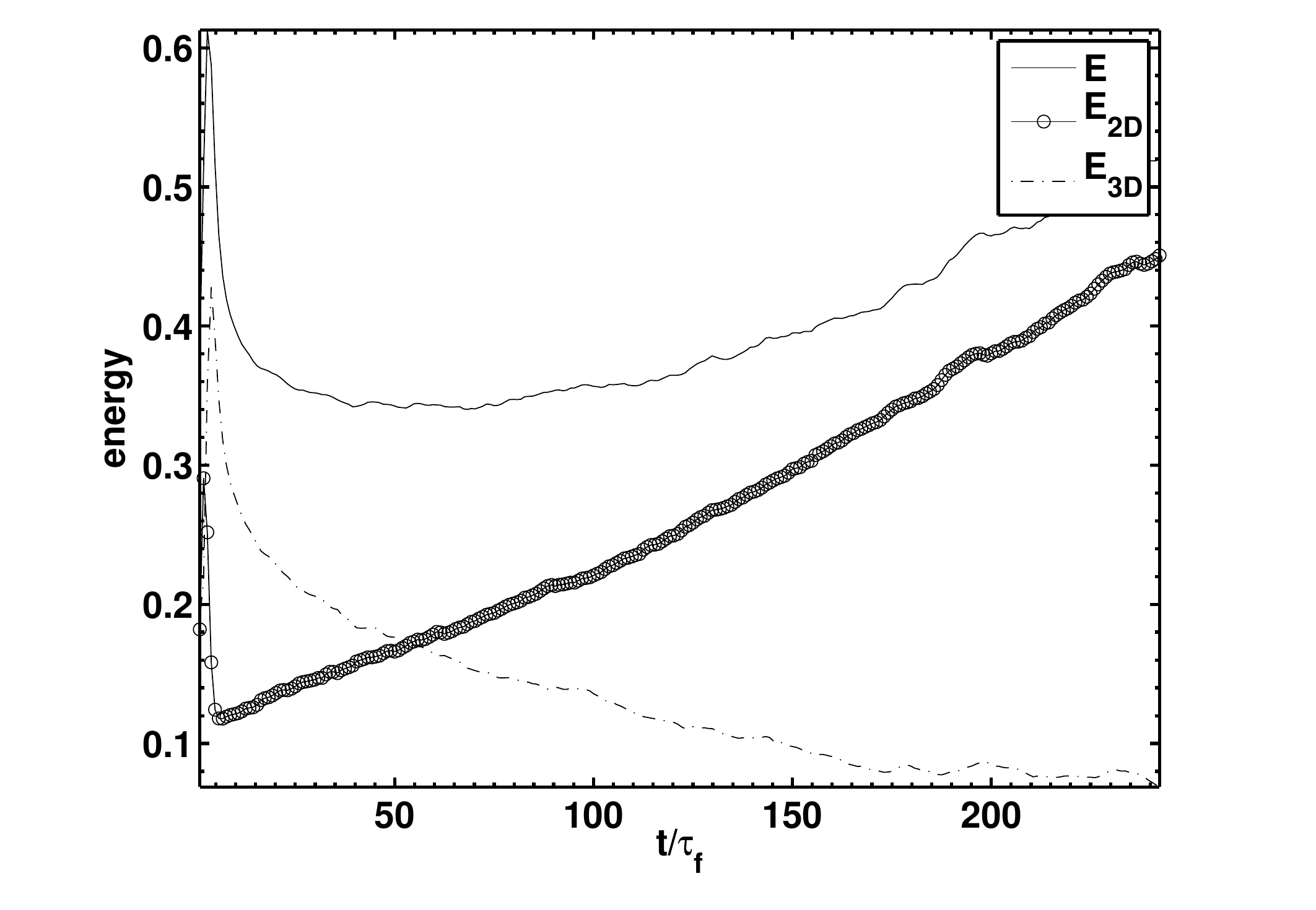}
\includegraphics[trim=0cm 6cm 0cm 7cm,clip=true,height=5.5cm,width=9.5cm]{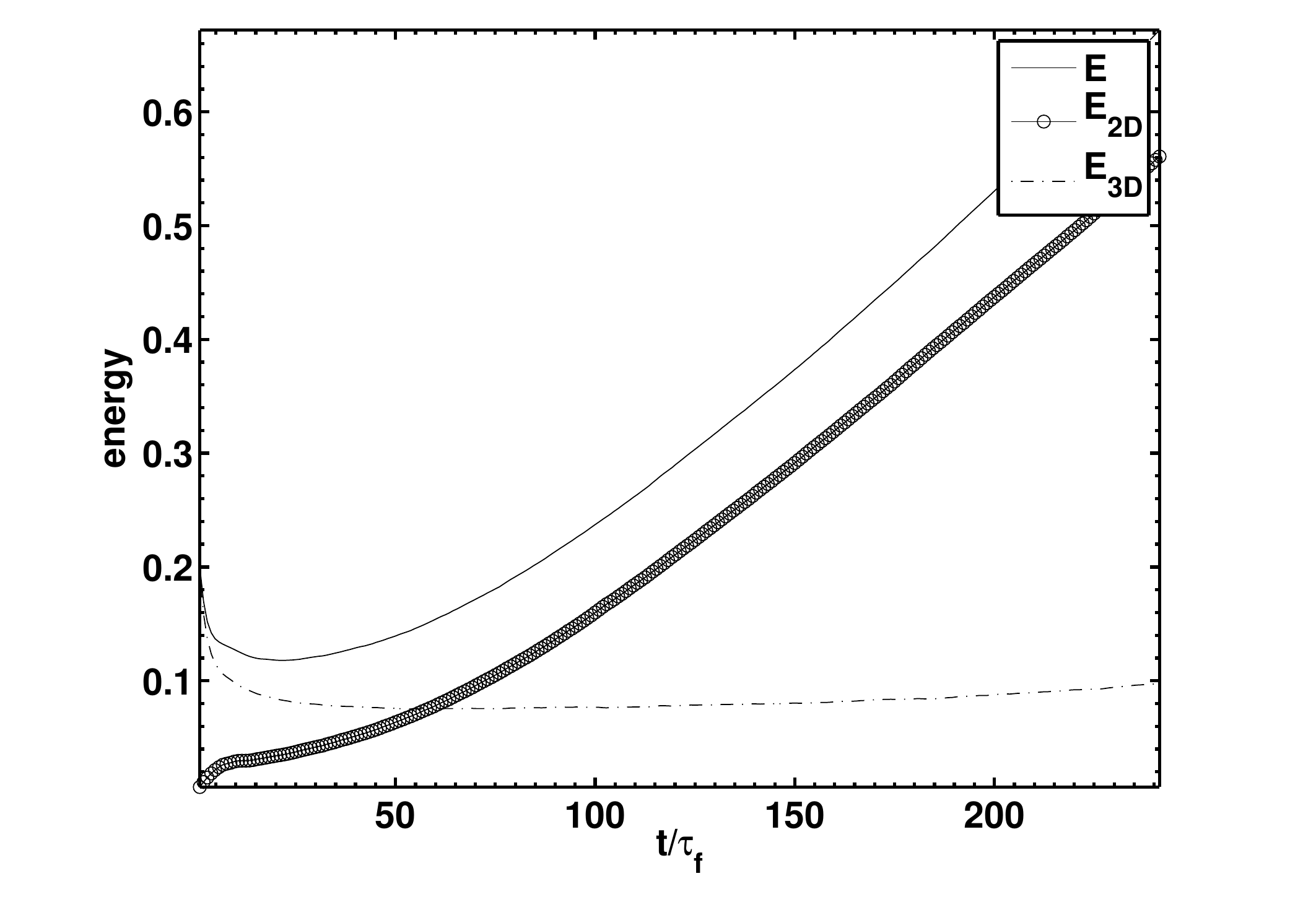}
\end{center}
\caption{Time evolution of $E$, $E_{2D}$, and $E_{3D}$ in runs TG, ABC and RND1 (from top to bottom).}
\label{fig:enerevol}
\end{figure}

\begin{figure}
\begin{center}
\includegraphics[trim=0cm 6cm 0cm 7cm,clip=true,height=5.2cm,width=9.5cm]{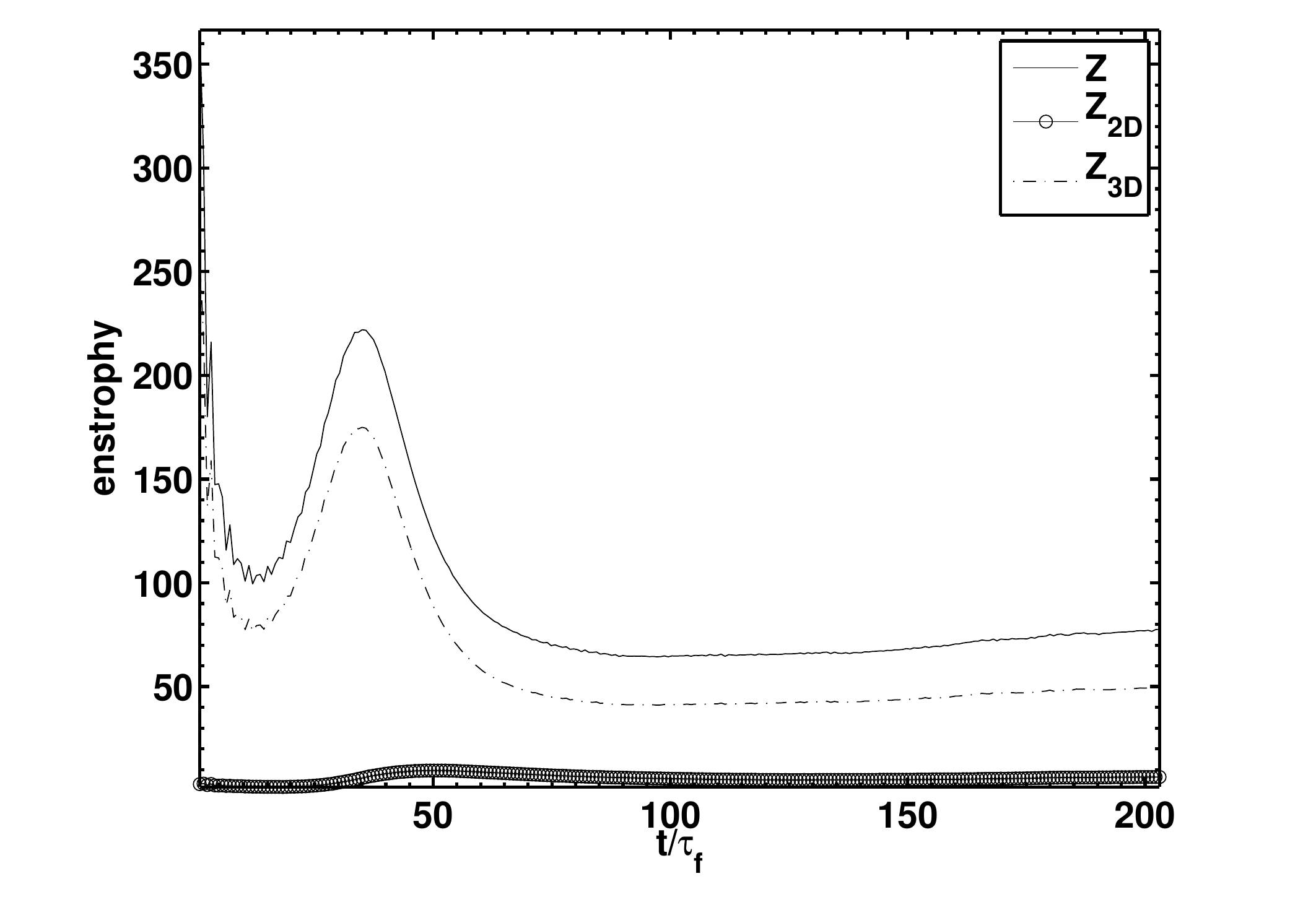}
\includegraphics[trim=0cm 6cm 0cm 7cm,clip=true,height=5.2cm,width=9.5cm]{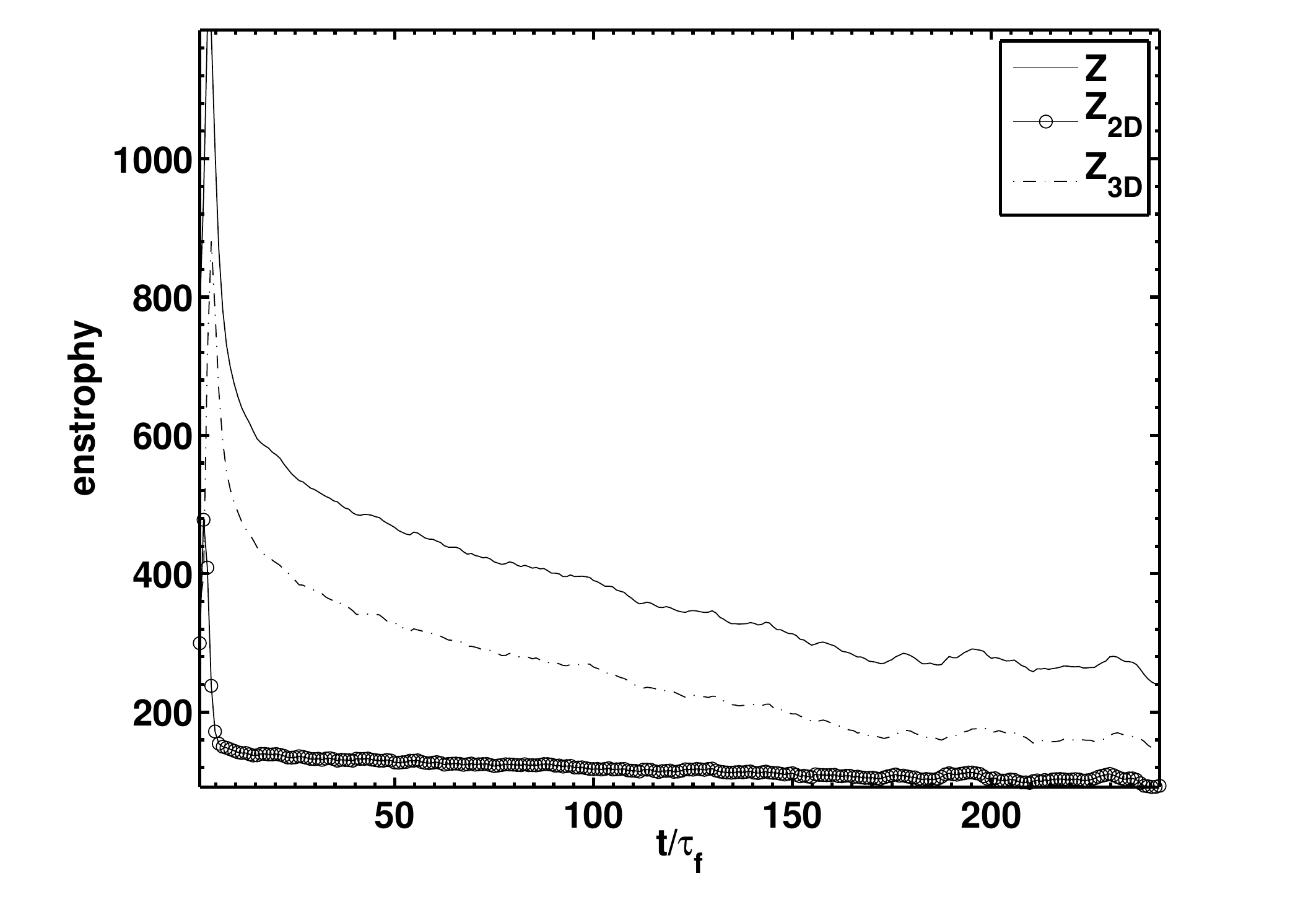}
\includegraphics[trim=0cm 6cm 0cm 7cm,clip=true,height=5.2cm,width=9.5cm]{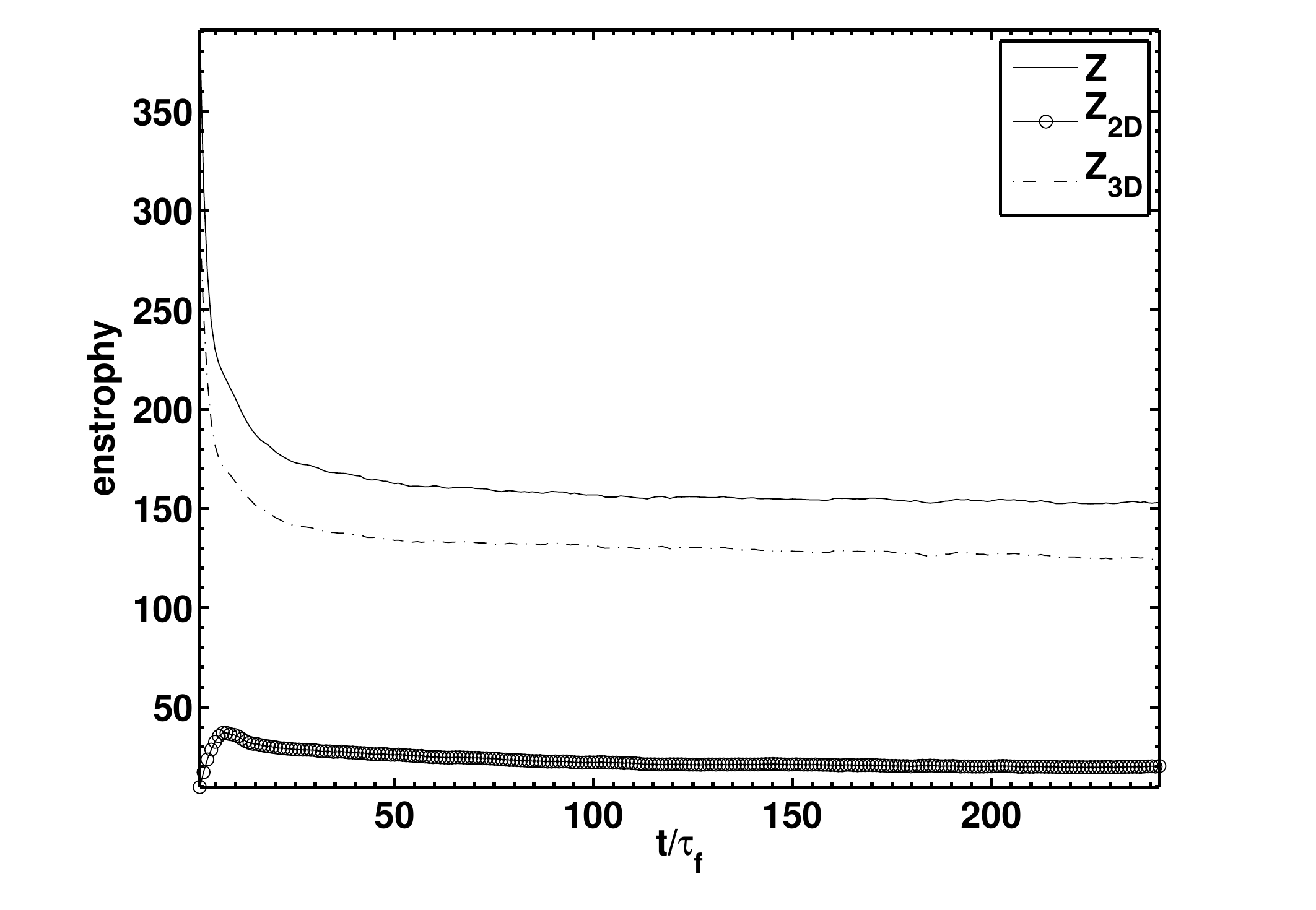}
\end{center}
\caption{Time evolution of the total enstrophy $Z$, enstrophy in 2D modes $Z_{2D}$, and enstrophy in 3D modes $Z_{3D}$ in runs TG, ABC and RND1 (from top to bottom).}
\label{fig:ensevol}
\end{figure}

The fluxes in Fig.~\ref{fig:flux1} do not discriminate between 2D and 3D modes, so although they confirm an inverse energy transfer, they are not enough to identify what modes are responsible for the large-scale pile up of energy. Figure \ref{fig:enerevol} shows the time evolution of $E$, $E_{2D}$, and $E_{3D}$ for several runs. In all cases, $E_{2D}$ grows monotonically in time after a short transient, thereby driving a growth of the total energy $E$. However, the energy in 3D modes, $E_{3D}$ can either slowly increase or decrease, depending on the forcing. On the other hand, the enstrophy in 2D modes:
\begin{equation}
Z_{2D} = \frac{1}{2}\sum_{{\bf k} \in V_k} |{\bf \omega}|^2 ,
\end{equation}
remains approximately constant once the inverse cascade starts, implying the small scales have reach{ed} a steady state. The enstrophy in 3D modes,
\begin{equation}
Z_{3D} = \frac{1}{2}\sum_{{\bf k} \in W_k} |{\bf \omega}|^2 ,
\end{equation}
as well as the total enstrophy $Z$, increases or decreases depending on the forcing function. Note that the transient regime at early times corresponds to a \textit{wave dominated} regime, and is longer for TG forcing as already reported in \cite{Mininni09}. Indeed, in the TG case, oscillations in $E_{3D}(t)$ and $Z_{3D}(t)$ with the frequency of inertial waves can be seen clearly before {$t/\tau_f\approx 20$.} 

\subsection{The effect of anisotropic energy injection}

\begin{figure}
\begin{center}
\includegraphics[trim=0cm 6cm 0cm 7cm,clip=true,height=5.2cm,width=9.5cm]{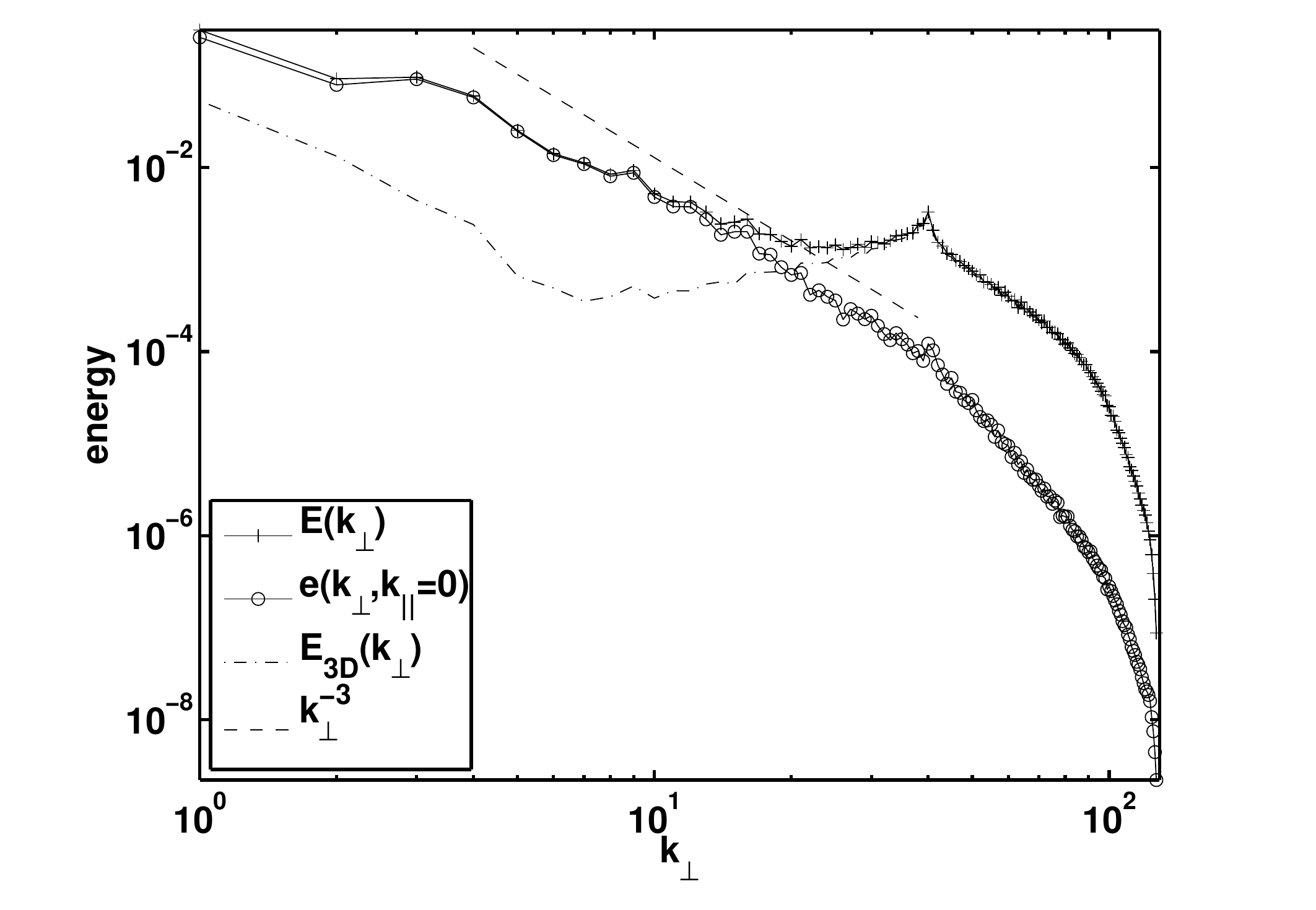}
\includegraphics[trim=0cm 6cm 0cm 7cm,clip=true,height=5.2cm,width=9.5cm]{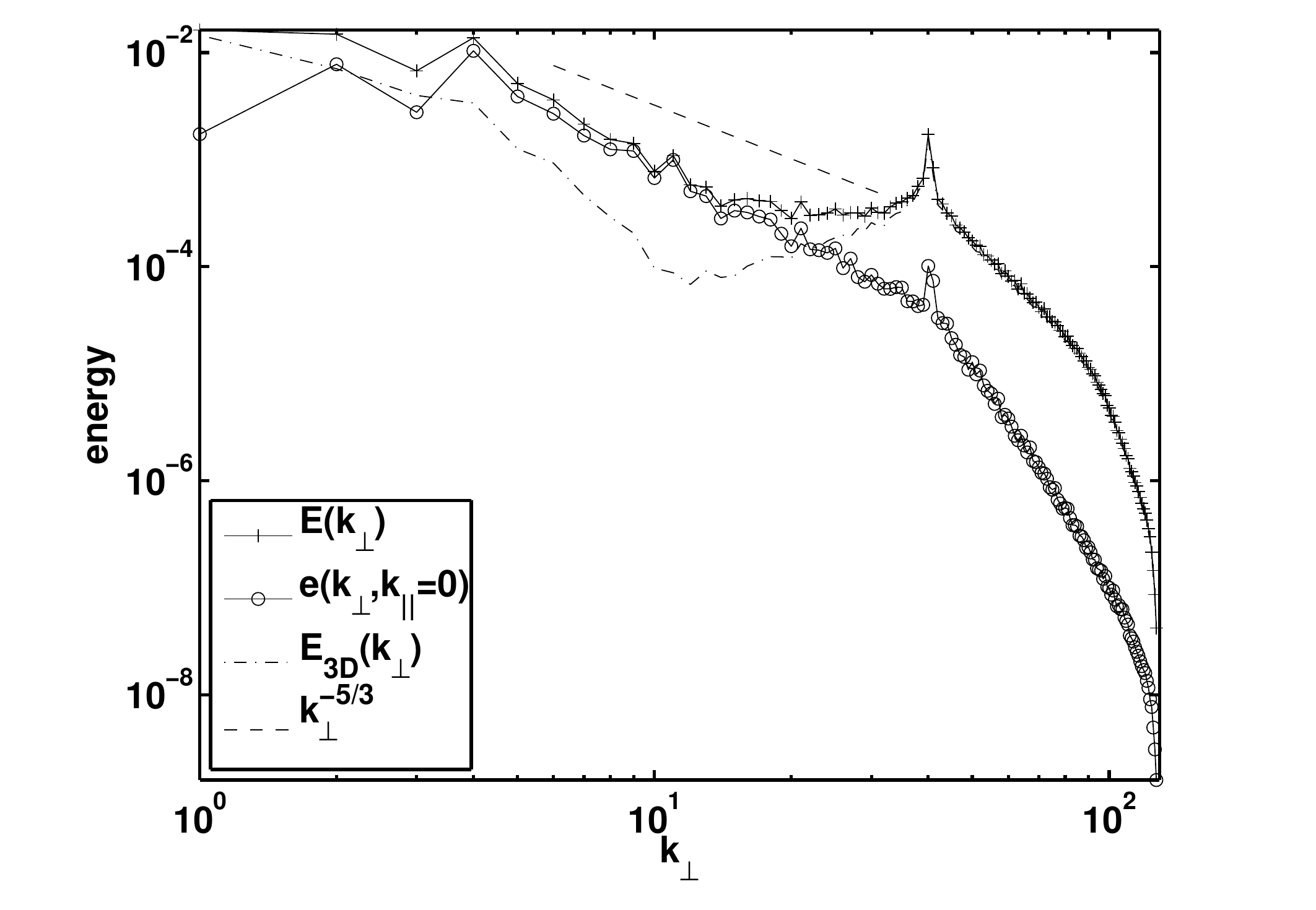}
\includegraphics[trim=0cm 6cm 0cm 7cm,clip=true,height=5.2cm,width=9.5cm]{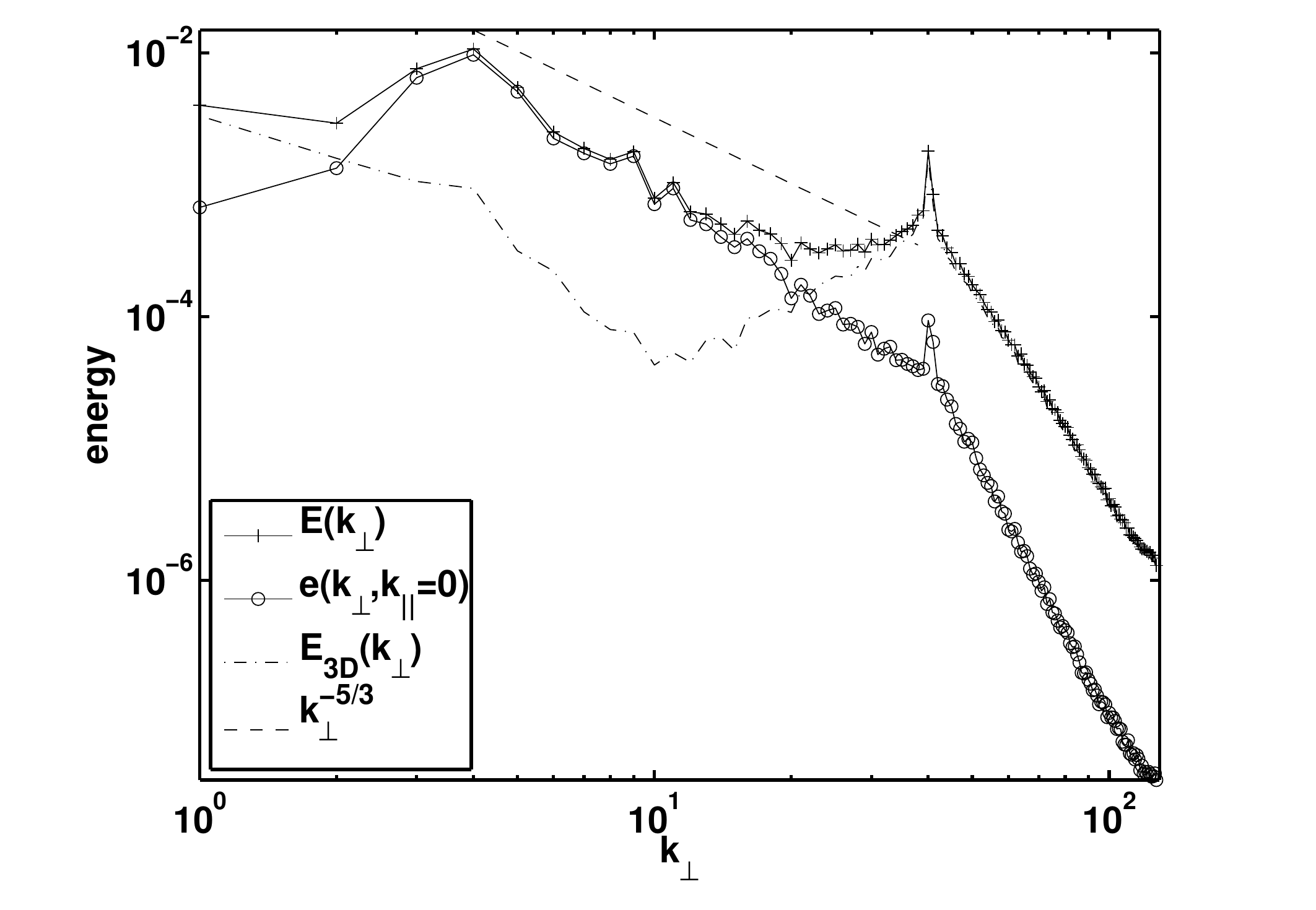}
\end{center}
\caption{Energy spectra at late times for runs ANI1 (top), ANI3 (middle) and ANI4 (bottom).}
\label{fig:specani6}
\end{figure}

Except for one case (ABC forcing), all simulations in the previous subsection seem to show an inverse cascade of 2D energy with a $k_\perp^{-3}$ scaling. What is the origin of the KKBL-like $\sim k_\perp^{-5/3}$ spectrum in the ABC run? Previous studies obtained KKBL scaling with elongated boxes \cite{Chen05} or when all triadic interactions between 2D and 3D modes were shut down \cite{Smith05} (which in fact corresponds to the case of KKBL phenomenology). However, in our case we used a box with fixed unit aspect ratio and with all triadic interactions and coupling between modes were accounted for in the simulations.

Helicity is the most conspicuous property of the flow generated by ABC forcing and it is known to affect the direct cascade range in rotating flows \cite{Mininni10,Pouquet10}. However, it is predominantly transferred to smaller scales as shown later (see also \cite{MinPou09}) and runs with isotropic {but} helical forcing (RND4) also show $\sim k_\perp^{-3}$ instead of KKBL scaling.

The difference between ABC forcing and the other forcing functions is that ${\bf f}_{\rm ABC}$ excites two 2D modes and one 3D mode in Fourier space, therefore effectively injecting more energy in 2D modes than in 3D modes. The other forcing functions considered in the previous subsection inject either more energy in 3D modes (RND), or energy only in a few 3D modes with no 2D injection (TG). Here we explore if anisotropic injection can be responsible for the different scaling laws observed in the $e(k_\perp,k_\parallel=0)$ spectrum by means of numerical simulations in which we control the anisotropy of the external forcing (runs ANI in table \ref{tab1}).

It should be pointed out that what we call here ``anisotropic injection'' for sake of brevity, is actually a more subtle effect associated with how much energy is directly injected in 2D modes compared to that into the 3D modes. Indeed, TG forcing is directionally anisotropic (in the sense that only a few modes in a spherical shell are excited in Fourier space), but it shows similar inverse cascade scaling as the RND runs, which have no directional anisotropy.

\begin{figure}
\begin{center}
\includegraphics[trim=0cm 6cm 0cm 7cm,clip=true,height=5.25cm,width=9.5cm]{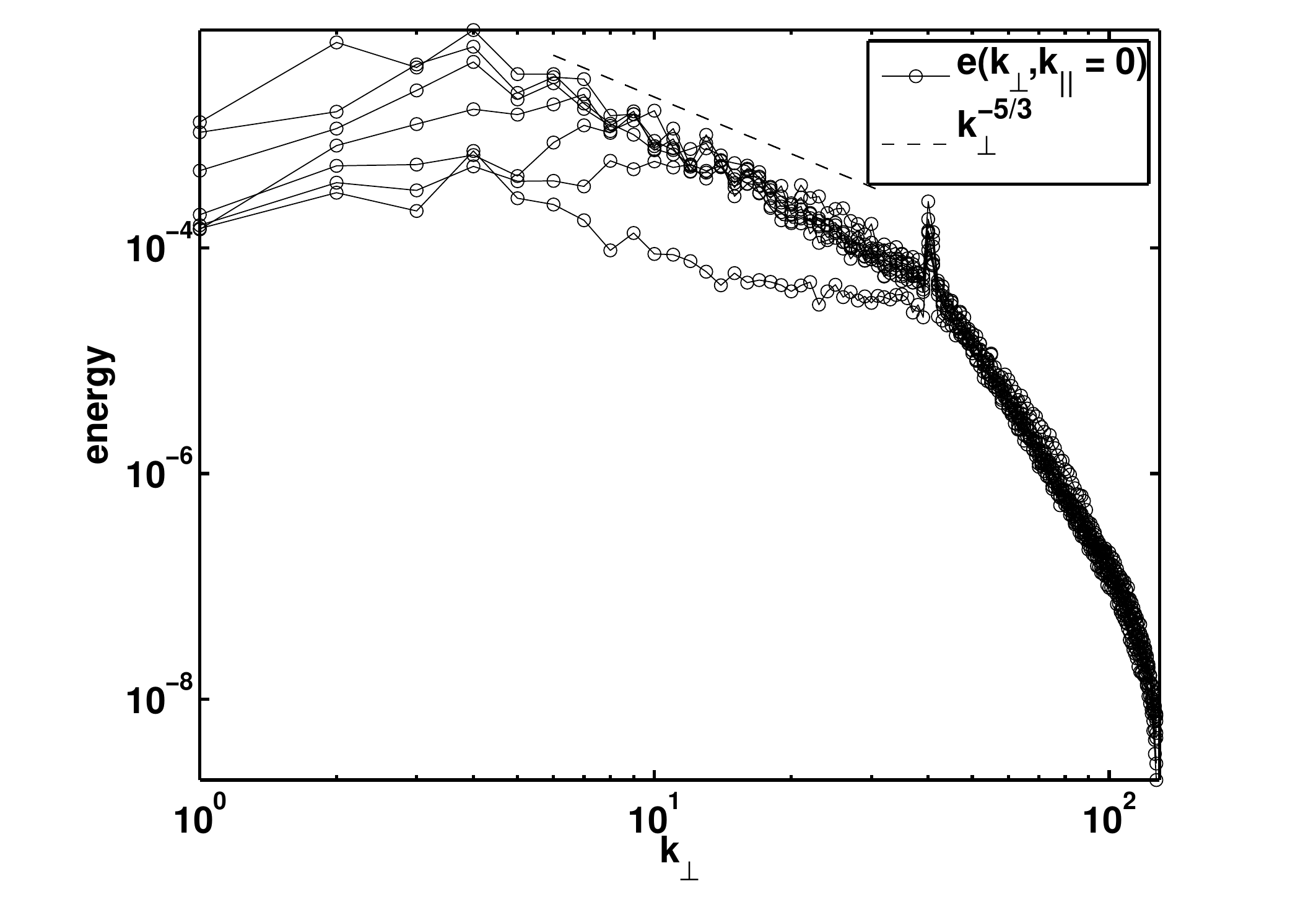}
\includegraphics[trim=0cm 6cm 0cm 7cm,clip=true,height=5.25cm,width=9.5cm]{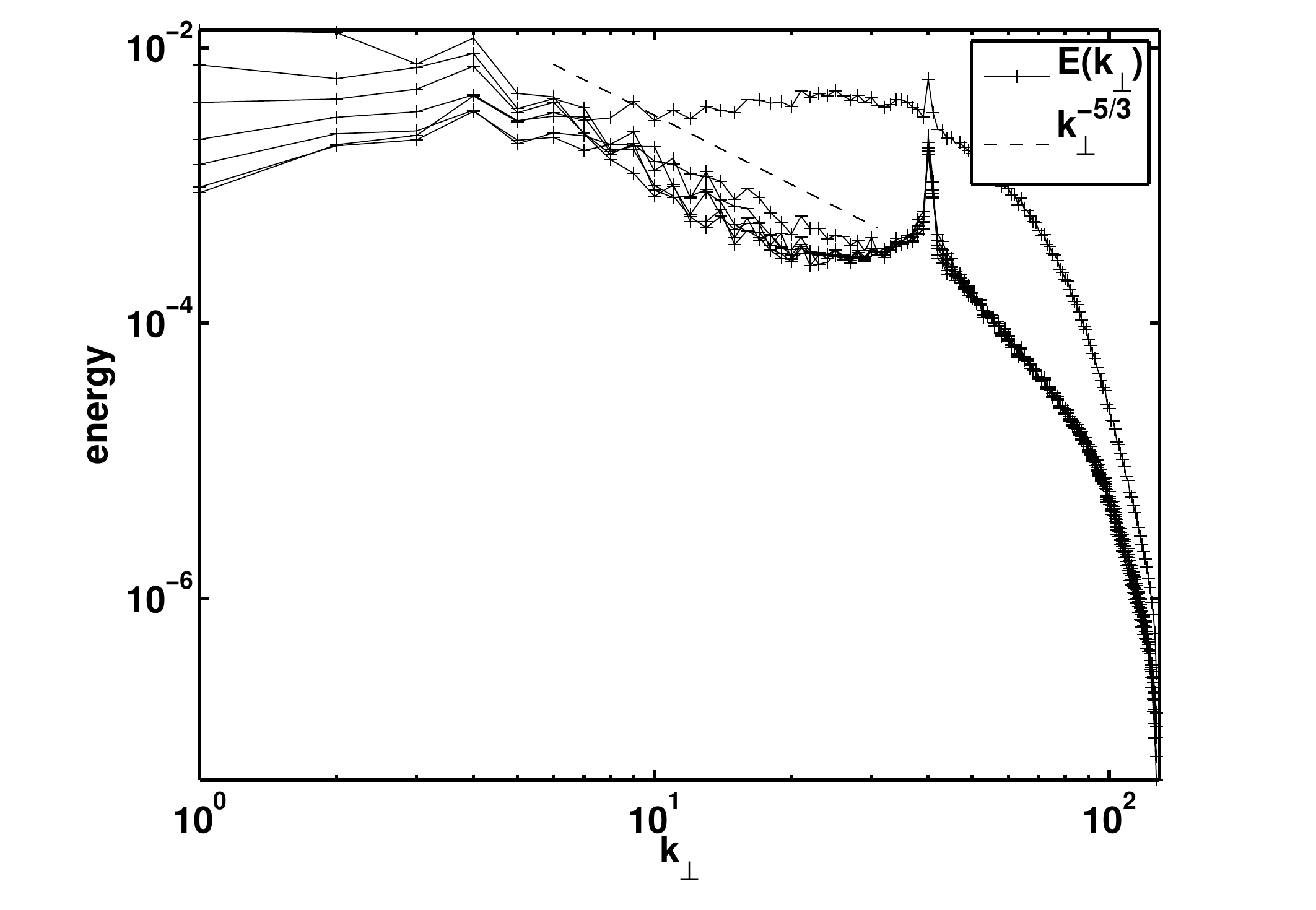}
\end{center}
\caption{Time evolution of the energy spectrum of 2D modes $e(k_\perp,k_\parallel=0)$ (top) and of the perpendicular spectrum $E(k_\perp)$ (bottom) in run ANI3. A $-5/3$ slope is indicated as a reference.}
\label{fig:specani}
\end{figure}

The energy spectra $e(k_\perp,k_\parallel=0)$, $E(k_\perp)$, and $E_{3D}(k_\perp)$ at late times for runs ANI1, ANI3, and ANI4 are shown in Fig.~\ref{fig:specani6}. The run ANI1 corresponds to a run with random forcing with anisotropic exponent $\beta=1$. Runs ANI3 and ANI4 have $\beta=3$, with zero and close to maximal helicity injection, respectively (see table \ref{tab1}). The spectra of ANI2 behave as ANI1 and are not shown.

The time evolution of $e(k_\perp,k_\parallel=0)$ and $E(k_\perp)$ for run ANI3, up to $250\tau_f$ turn-over times at intervals of $35.7\tau_f$ turn-over times, is shown in Fig.~\ref{fig:specani}. A clear build up of a $\sim k_\perp^{-5/3}$ spectrum can be observed in the runs with anisotropic forcing irrespective of the amount of helicity in the flow. 

\subsection{Helicity at large scales}

\begin{figure}
\begin{center}
\includegraphics[trim=0cm 6cm 0cm 7cm,clip=true,height=5.25cm,width=9.5cm]{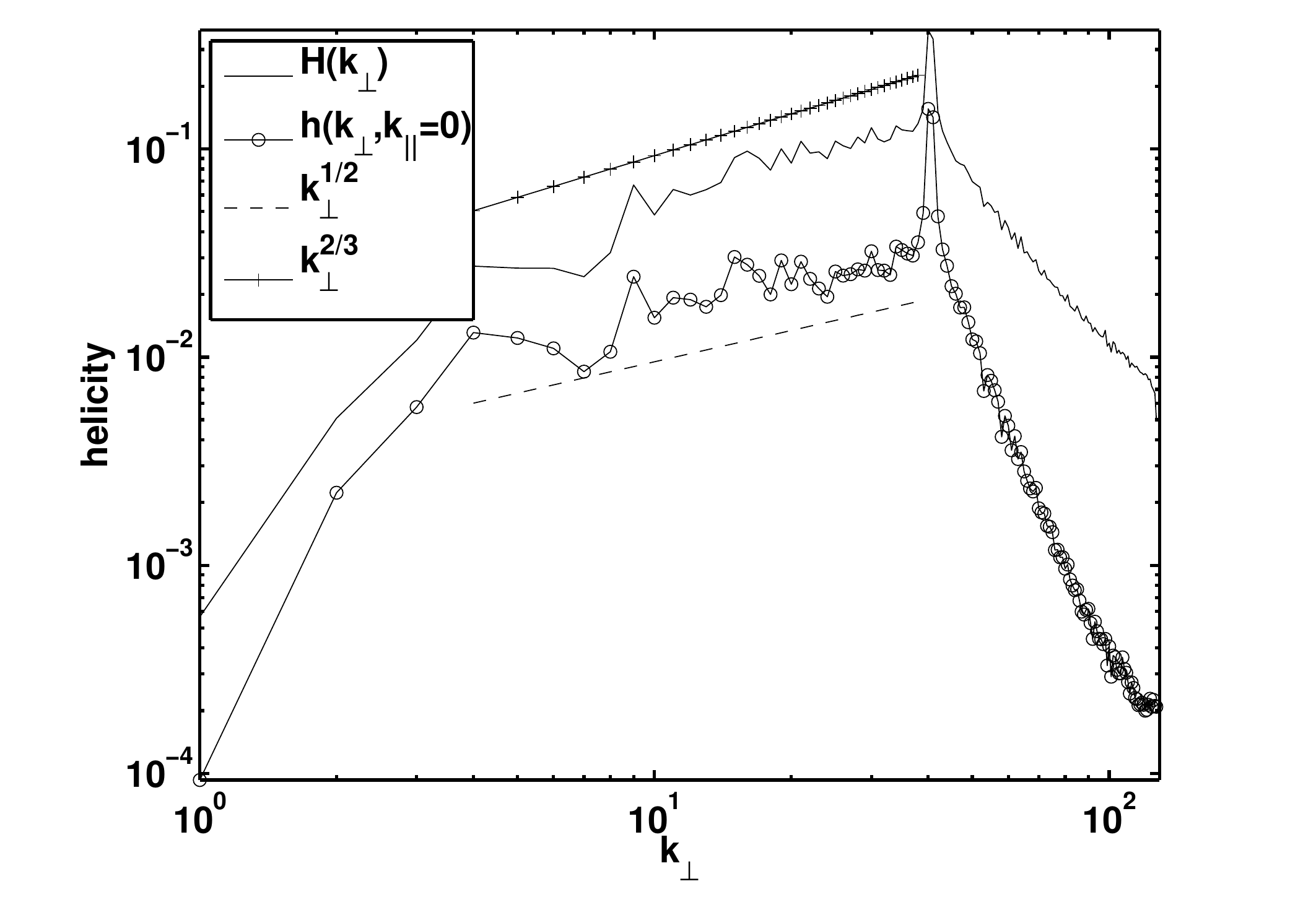}
\includegraphics[trim=0cm 6cm 0cm 7cm,clip=true,height=5.25cm,width=9.5cm]{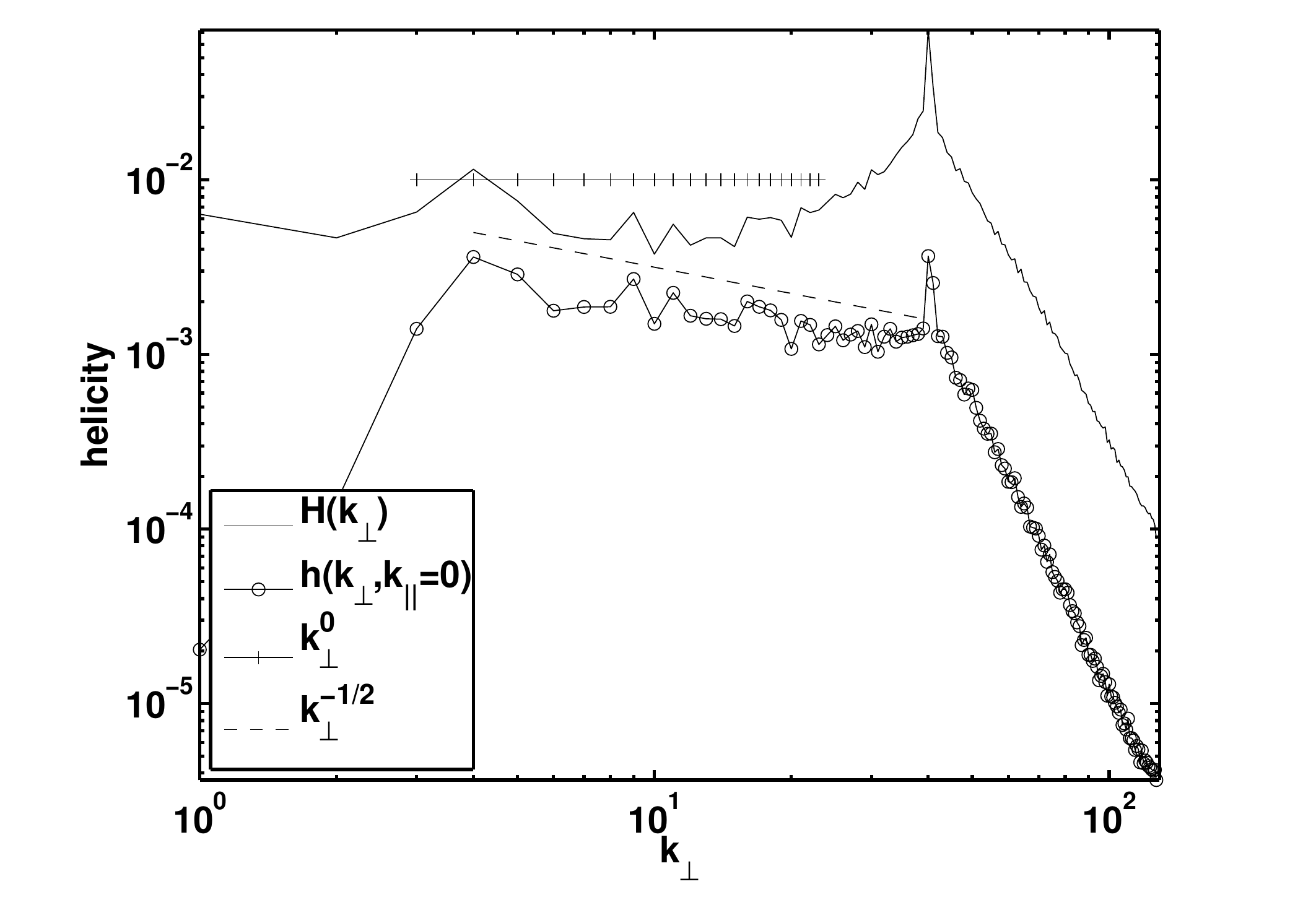}
\end{center}
\caption{Helicity spectra $H(k_\perp)$ and $h(k_\perp,k_\parallel=0)$ (i.e., the spectrum of helicity contained in purely horizontal motions) for runs ABC (above) and ANI4 (below). Slopes are indicated as a reference.}
\label{fig:helispec} \end{figure}

Fig.~\ref{fig:helispec} shows the helicity spectra $H(k_\perp)$ and $h(k_\perp,k_\parallel=0)$ for the runs with helical forcing ABC and ANI4. There is no significant large scale growth of helicity and the relative helicity remains negligibly small at low wavenumbers, $\rho_H(k) = H(k)/k E(k) \to 0$ for $k \to 0$ (also $\rho_H(k_\perp) \to 0$ for $k_\perp \to 0$). This is consistent with the observation above that helicity does not seem to affect the energy scaling in the inverse cascade range, as has also been shown in previous studies of helicity cascading to smaller scales in rotating flows \cite{MinPou09}.

\subsection{Coupling and fluxes between slow and fast modes}

\begin{figure}
\begin{center}
\includegraphics[trim=0cm 6.5cm 0cm 6cm,clip=true,height=6cm,width=9.5cm]{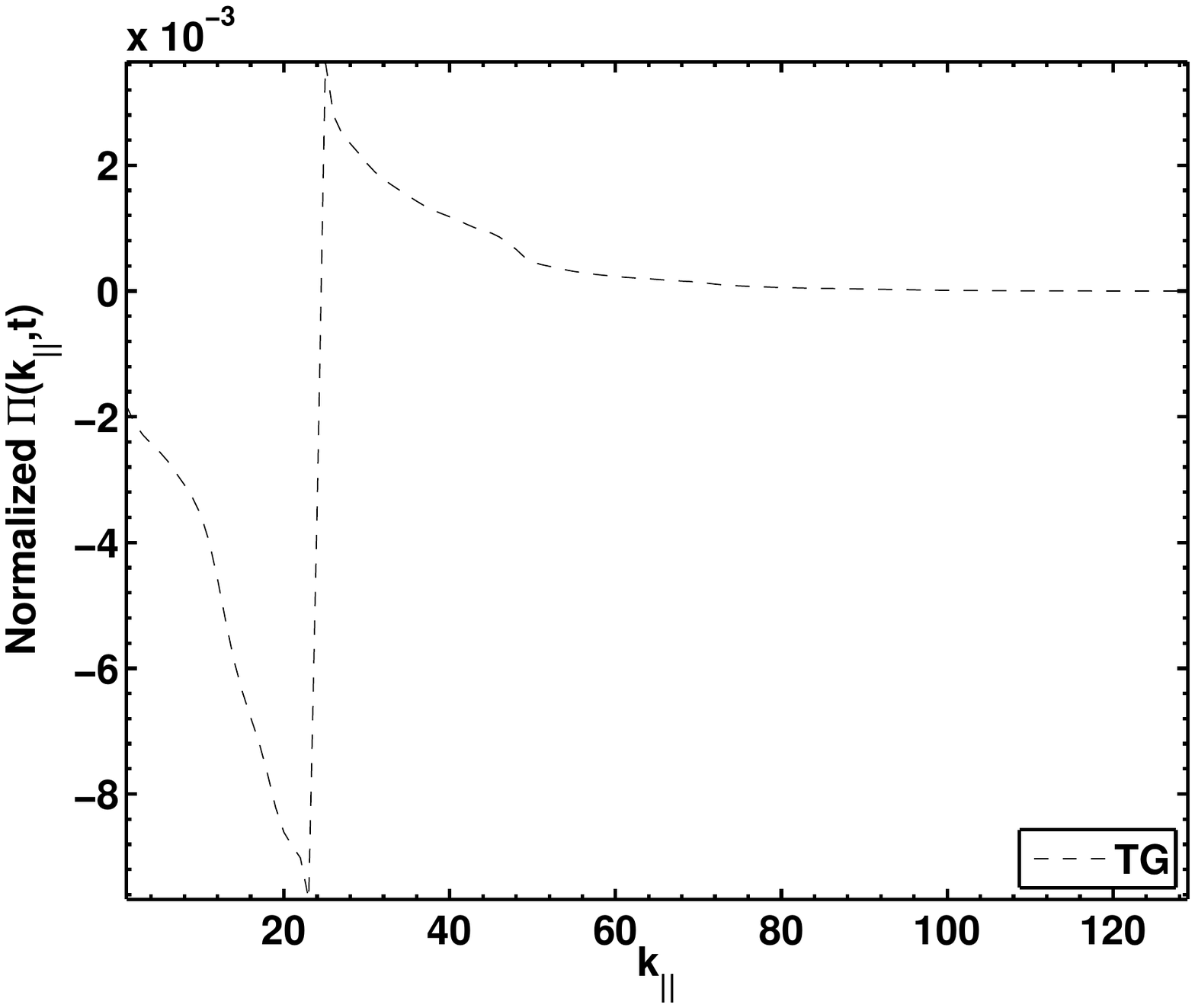}
\includegraphics[trim=0cm 6.5cm 0cm 6cm,clip=true,height=6cm,width=9.5cm]{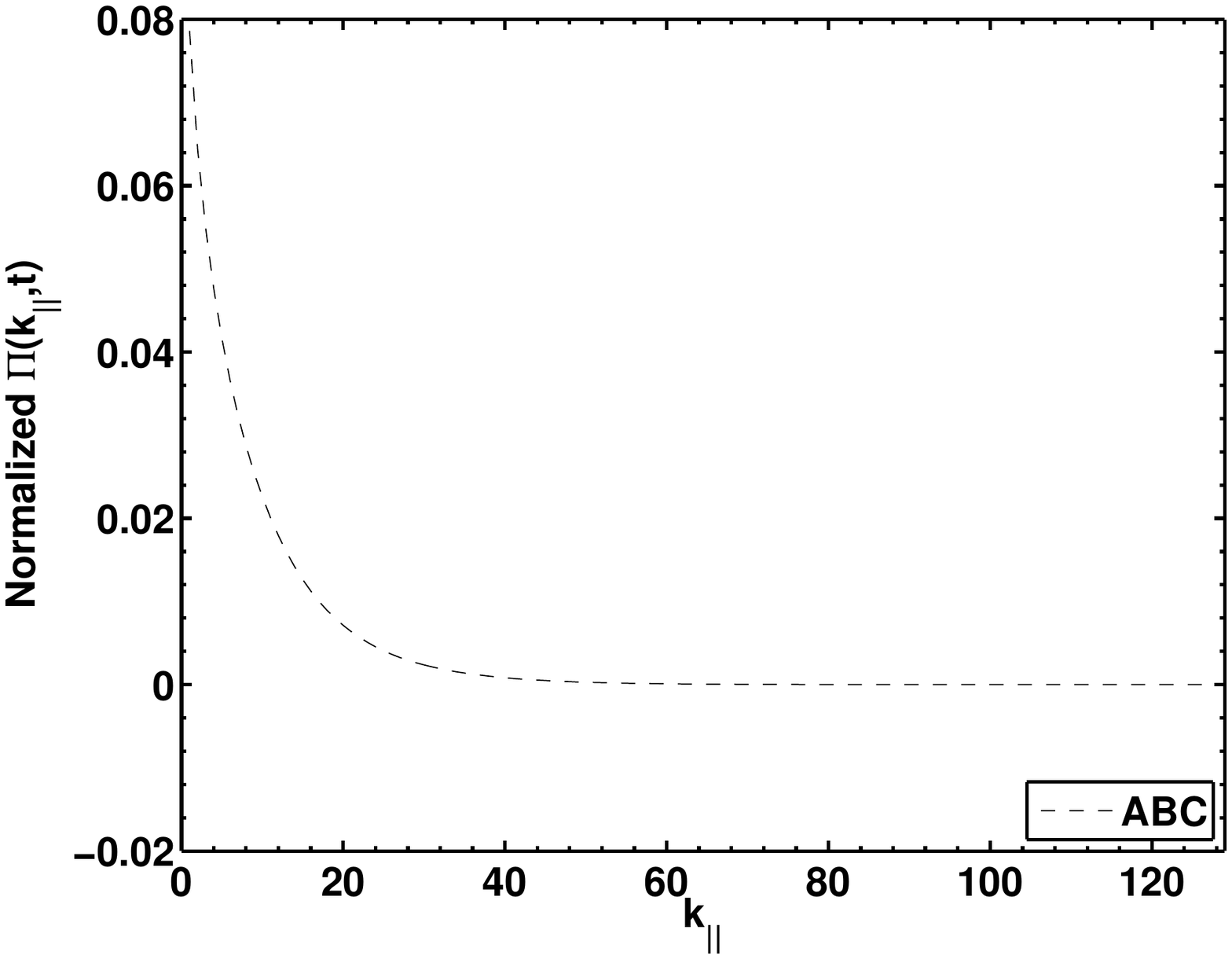}
\includegraphics[trim=0cm 6.5cm 0cm 6cm,clip=true,height=6cm,width=9.5cm]{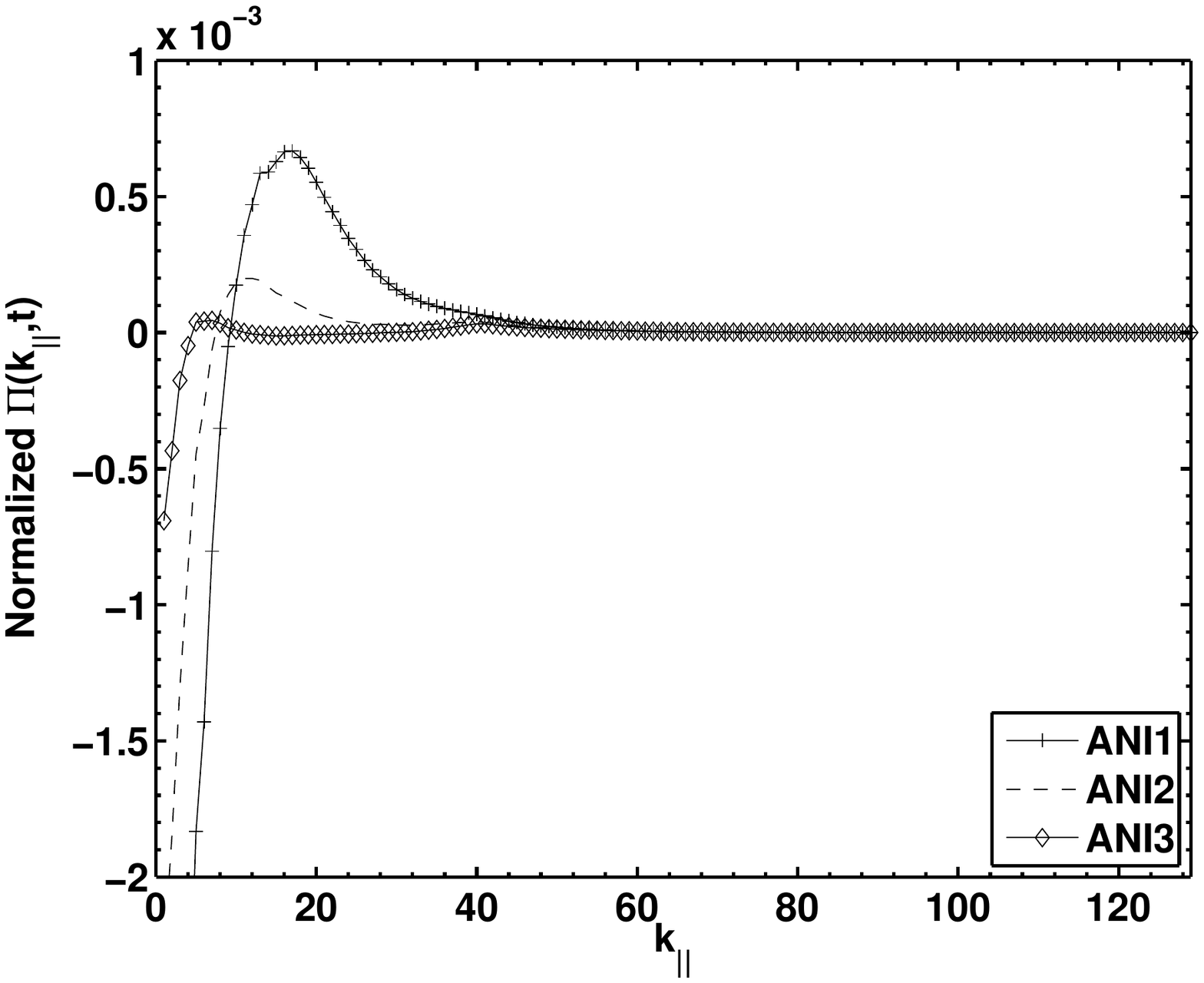}\\
\end{center}
\caption{{$\Pi(k_\parallel)$ for runs TG, ABC, ANI1, ANI2 and ANI3 (top to bottom).}}
\label{fig:aniflux}
\end{figure}

\begin{table*}
\caption{Amplitude of the terms in Eq.~\eqref{EnFlx3D}. The time derivative $dE_{3D}/dt$ was obtained using centered finite differences from the data. $\Pi(k_\parallel=0)$ represents energy per unit of time transferred from 2D to 3D modes, and $\epsilon_{3D}$ is the power injected in the 3D modes. $\Pi_{3D}^{\rm l.h.s.}$ is the flux of energy in the 3D modes estimated from Eq.~(\ref{eq:3Dlhs}), $\Pi_{3D}^{\rm est.}$ is estimated from Eq.~\eqref{eq:estimation}, and $2\nu\int k^2Z_{3D}(k)\, dk$ is an estimation based on the energy dissipation rate. }
\begin{ruledtabular}  \begin{tabular}{ccccccc}
Run  &  $dE_{3D}/dt$   & $\Pi(k_\parallel=0)$& $\epsilon_{3D}$ & $\Pi_{3D}^{\rm l.h.s.}$& $\Pi_{3D}^{\rm est.}$& $2\nu\int k^2E_{3D}(k) dk$ \\
\hline
TG   & $1.0\times10^{-4}$&  $-2.0\times10^{-3}$ & $3.0\times10^{-2}$& $2.8\times10^{-2}$     & $1.0\times10^{-2}$    &  $1.0\times10^{-2}$          \\
RND1 & $1.0\times10^{-4}$&  $-6.8\times10^{-3}$ & $4.6\times10^{-2}$& $3.9\times10^{-2}$     & $2.0\times10^{-2}$    &  $2.0\times10^{-2}$          \\
RND4 & $4.0\times10^{-5}$&  $-6.3\times10^{-3}$ & $4.6\times10^{-2}$& $3.9\times10^{-2}$     & $2.0\times10^{-2}$    &  $3.0\times10^{-2}$          \\
ANI1 & $1.0\times10^{-4}$&  $-5.2\times10^{-3}$ & {$8.9\times10^{-3}$}& {$3.6\times10^{-3}$}     & $4.0\times10^{-2}$    &  $4.0\times10^{-2}$          \\
ANI2 & $1.0\times10^{-4}$&  $-2.3\times10^{-3}$ & {$9.3\times10^{-3}$}& {$6.9\times10^{-3}$}     & $2.0\times10^{-2}$    &  $1.0\times10^{-2}$          \\
ANI3 & $1.0\times10^{-4}$&  $-6.9\times10^{-4}$ & {$6.5\times10^{-3}$}& {$5.7\times10^{-3}$}     & $2.0\times10^{-3}$    &  $5.0\times10^{-3}$          \\
ANI4 & $3.0\times10^{-5}$&  $-6.0\times10^{-4}$ & {$5.4\times10^{-3}$}& {$4.7\times10^{-3}$}     & $2.0\times10^{-3}$    &  $4.0\times10^{-3}$          \\
ABC  &$-2.0\times10^{-4}$&  $8.3\times10^{-2}$  & $2.0\times10^{-2}$& $1.0\times10^{-1}$     & $3.0\times10^{-2}$    &  $3.0\times10^{-2}$
\end{tabular} \end{ruledtabular}
\label{tab2} \end{table*}

\begin{table}
\caption{Amplitude of the terms in Eq.~(\ref{EnFlx2D}). The time derivative $dE_{2D}/dt$ was obtained using centered finite differences, $\epsilon_{2D}$ is the power injected in the 2D modes, and $\Pi_{2D}^{\rm l.h.s.}$ is the flux of energy in 2D modes estimated from Eq.~(\ref{eq:2Dlhs}).}
\begin{ruledtabular} \begin{tabular}{cccc}
Run  &  $dE_{2D}/dt$   & $\epsilon_{2D}$ & $\Pi_{2D}^{\rm l.h.s.}$ \\
\hline
TG   &  {$4.0\times10^{-4}$}        &  {$1.0 \times 10^{-10}$}            & {$1.6 \times 10^{-3}$}              \\ 
RND1 &  {$1.0\times10^{-3}$}        &  {$1.3 \times 10^{-3}$}      &   {$7.1\times10^{-3}$}       \\ 
RND4 &  {$2.0\times10^{-3}$}        &  {$1.5 \times 10^{-3}$}   &  {$5.8\times10^{-3}$}             \\ 
ANI1 &  {$1.0\times10^{-3}$}        &   {$1.1 \times 10^{-3}$}       &  {$5.3\times10^{-3}$}           \\ 
ANI2 &  {$4.0\times10^{-4}$}              &  {$7.0 \times 10^{-4}$}          &   {$2.6\times10^{-3}$}   \\ 
ANI3 &  {$7.0\times10^{-5}$}              &    {$5.0 \times 10^{-4}$}         &     {$1.1\times10^{-3}$}    \\ 
ANI4 &  {$8.0\times10^{-5}$}   &  {$6.4 \times 10^{-4}$}  &   {$1.1\times10^{-3}$} \\ 
ABC  & {$5.0\times10^{-4}$}     &  {$7.0 \times 10^{-2}$}        &    {$-1.3\times10^{-2}$}           \\ 
\end{tabular} \end{ruledtabular}
\label{tab3} \end{table}  

\begin{figure}
\begin{center}
\includegraphics[height=6.1cm,width=9.5cm]{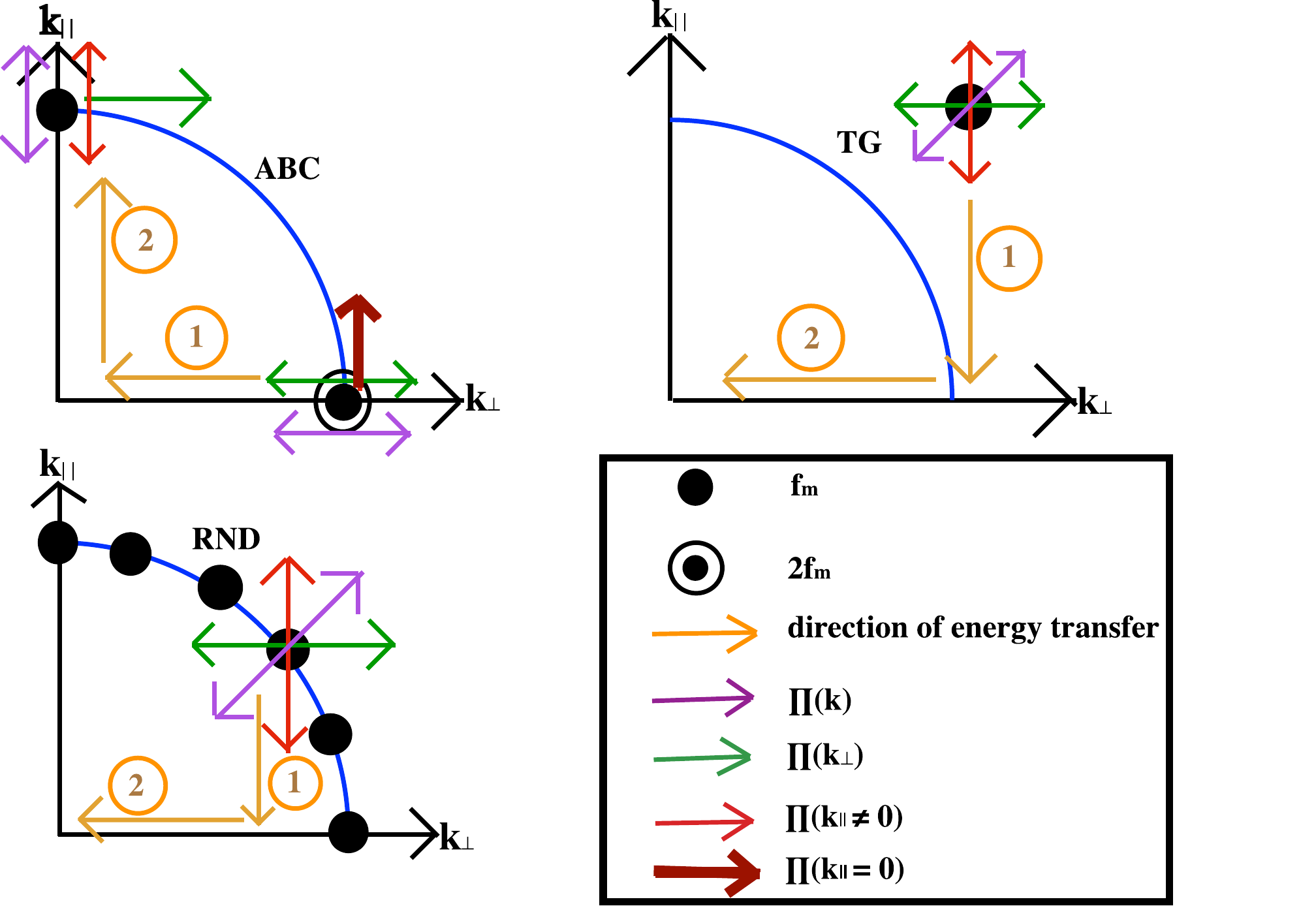}
\end{center}
\caption{A schematic depiction of the direction of transfer of energy (and corresponding fluxes) in the case of forced rotating turbulence. Here $f_m$ is the normalized unit amplitude of forcing in Fourier space. The black dots indicate the modes that are directly excited by the different forcing functions, with $2f_m$ indicating twice the energy injected in that mode compared to the energy injected into other modes. Directions of the arrows are based on the analysis of data from tables \ref{tab2}, \ref{tab3} and Figs.~\ref{fig:specevol1}, \ref{fig:enerevol}. }
\label{fig:en_tr_sch} 
\end{figure}

How does the amount of energy injected into the 2D and the 3D modes affect the inverse cascade? Fig.~\ref{fig:aniflux} shows the flux $\Pi(k_\parallel)$ for runs TG, ABC, ANI1, ANI2, and ANI3 specified in table \ref{tab1}. In most runs, the flux is negative for small values of $k_\parallel$ (indicating that energy goes from the 3D modes towards the 2D modes for larger scales), and positive for large values of $k_\parallel$ (indicating that energy goes away from the 2D modes for smaller scales). Note that as more energy is injected into the 2D modes (e.g., as $\beta$ is increased in the ANI runs), the wavenumber at which the fluxes change sign moves towards $k_\parallel=0$, and for the ABC flow the flux $\Pi(k_\parallel)$ is positive everywhere (i.e., energy goes from the 2D modes to the 3D modes
{at all scales).} As a reference, a schematic representation of the excited modes in axisymmetric Fourier space $(k_\perp,k_\parallel)$ is shown in Fig.~\ref{fig:en_tr_sch}.

The above observations imply that the ABC flow corresponds to the limiting case in which most of the energy is injected into the 2D modes and as a result of the imbalance, an excess of energy ``leaks'' from the 2D modes to the 3D modes. This can be verified by computing each term in the energy balance Eqs.~\eqref{EnFlx3D} and \eqref{EnFlx2D} (see tables \ref{tab2} and \ref{tab3}). The flux of 3D and 2D energy can be estimated from {Eqs.~\eqref{EnFlx3D} and \eqref{EnFlx2D}} as follows. Since all terms in these equations, with the exception of $\Pi_{2D}$ and {$\Pi_{3D}$},  are known, the equations can be re-written as:
\begin{equation}
\Pi_{3D}^\textrm{l.h.s} = \Pi(k_{||}=0,t) + \epsilon_{3D} -d_t E_{3D} ,
\label{eq:3Dlhs}
\end{equation}
and
\label{eq:2Dlhs}
\begin{equation}
\Pi_{2D}^\textrm{l.h.s} = -\Pi(k_{||}=0,t) + \epsilon_{2D} - d_t E_{2D}.
\label{eq:2Dlhs}
\end{equation}
The superscript ``l.h.s.''~here and in the table indicates that the fluxes are obtained by solving for the l.h.s.~of the balance equations above. Another way of estimating $\Pi_{3D}$ {is based on geometrical consideration of the fluxes in spectral space (see Fig.~\ref{fig:en_tr_sch})}, 
\begin{equation}
\Pi_{3D}^\textrm{est} =
\begin{cases}
 max_{k_f \le k \le k_{f+1}} \{\Pi(k,t)\} & \text{if ${\bf f} = {\bf f_{ABC}}$}\\
 max_{\forall k} \{\Pi(k) - \Pi(k_{\perp})\} &\text{if ${\bf f} = {\bf f_{TG}},{\bf f_{RND}}$}.
\end{cases}
\label{eq:estimation}
\end{equation}
Finally, {an alternative interpretation of $\Pi_{3D}$ is that} it transfers energy to the 3D modes with larger wavenumbers where it is eventually dissipated, and hence it balances the injection term with dissipation of energy per unit of time. Hence, it can also be approximated by $2\nu \int_{0}^{k_{max}}k^2E_{3D}(k)dk$. The two estimates have been found to be of the same order of magnitude in all the runs as shown in table \ref{tab2}. The fact that the three estimates are positive and of the same order indicate that energy in the 3D modes does not only go to the slow manifold but goes to smaller scales where it dissipates. The fact that $\Pi_{2D}$ is negative in the ABC run (see table \ref{tab3}) is evidence of an inverse cascade of energy in the slow manifold once the energy from the 3D modes is transferred to the 2D modes. For the other runs, even though $\Pi_{2D}$ is positive and small in magnitude, it merely implies that more energy cascades to the smaller scales than to the larger scales. It is important to note that the positiveness of $\Pi_{2D}$ hints at positive eddy viscosity and the possibility of the inverse cascade of energy in the slow manifold cannot be ruled out. In fact, the nature of the evolution of the $e(k_{\perp},k_{||}=0)$ spectra over time, as shown in Fig.~\ref{fig:specevol1}, is evidence of occurrence of the inverse cascade of energy in the slow manifold. It may also be worth pointing out that with increasing $\beta$ (anisotropy), $\Pi_{2D}$, for the ANI runs, become less positive and seems to approach the nature of the energy cascade exhibited by the ABC run (see table \ref{tab3}). 

The picture that emerges for the fluxes from the values in tables \ref{tab2} and \ref{tab3} is illustrated schematically in Fig.~\ref{fig:en_tr_sch}. For isotropic forcing, a fraction of the energy injected into the 3D modes is transferred to the slow manifold and the remaining 3D energy is transferred to the 3D modes with larger wavenumbers.  

Recently, \cite{Bourouiba2011} did a detailed study of energy transfers in forced rotating turbulence and concluded that the former transfer, from the 3D to the 2D modes, is non-local. The results in \cite{Bourouiba2011} (with simulations forced at smaller wavenumbers than in our case) are consistent with our results, except that they attribute the development of the $\sim k^{-3}$ spectrum in the 2D modes to a direct cascade of enstrophy once the 3D modes inject energy directly into the 2D modes with the smallest wavenumbers. Although our analysis cannot disprove this conjecture, the amplitude of the terms in tables \ref{tab2} and \ref{tab3} and the nature of the evolution of the 2D energy spectra in Fig.~\ref{fig:specevol1} hints at a likely inverse transfer of energy in the slow manifold.

For anisotropic and ABC forcing the picture changes. As more energy is injected directly into the 2D modes by the forcing, the flux of energy from the 2D to the 3D modes, $\Pi(k_\parallel=0)$ in table \ref{tab2}, increases (from larger negative values to smaller negative values) and eventually reverses sign becoming positive. In the ABC flow (and likely for other flows with very high anisotropic forcing), the energy injected directly into the 2D modes undergoes an inverse cascade in the slow manifold, and later the excess of energy in these modes relative to the 3D modes ``leaks'' energy into the 3D modes at large scales (see Fig.~\ref{fig:en_tr_sch}). As the effect of the 3D modes over the 2D modes is less relevant in these runs, the runs with either small or positive $\Pi(k_\parallel=0)$ display $\sim k_\perp^{-5/3}$ scaling.

It is also interesting to point out that the 2D and the 3D modes are not necessarily decoupled, as can be seen, e.g., by comparing the ratios $\Pi(k_\parallel=0)/\Pi_{3D}^{\rm l.h.s.}$ and $\Pi(k_\parallel=0)/\Pi_{2D}^{\rm l.h.s.}$. The relevant quantity for either a $\sim k_\perp^{-3}$ or $\sim k_\perp^{-5/3}$ scaling in the inverse cascade energy spectrum is $\Pi(k_\parallel=0)$. Indeed, based on the previous discussion, if little energy goes into the 2D modes from the 3D modes, or if energy goes from the 2D modes into the 3D modes with the 2D modes being the most energetic and dominating the dynamics, we can assume that the cascade in the slow manifold is dominated by the turnover time $\tau_\perp \sim l_\perp/u_\perp$ (where $l_\perp$ is a characteristic length in the slow manifold and $u_\perp$ the 2D r.m.s.~velocity at that length). With only one relevant timescale, KKBL phenomenology tells us that the energy flux goes as:
\begin{equation}
\Pi_{2D} \sim \frac{u_\perp^2}{\tau_\perp} \sim \frac{u_\perp^3}{l_\perp} ,
\end{equation}
which results in a $\sim k_\perp^{-5/3}$ scaling law. On the other hand, if energy goes from the 3D modes to the 2D modes, interactions with the 3D modes cannot be neglected. Besides the slow turnover time $\tau_\perp$, we have to consider now the 3D turnover time and the timescale associated with the fast waves, $\tau_\Omega \sim \frac{1}{\Omega}$. There is no unique dimensional solution in this case {but we can borrow from the phenomenology developed by Kraichnan \cite{Kraichnan65} for magnetohydrodynamic  (MHD) turbulence where the effect of the waves modulates the dominant time-scale of the flow. This phenomenology has been successfully extended to rotating flows \cite{dubrulle, Zhou95}, including in the helical case \cite{Mininni10}. It states that in the presence of waves, the nonlinear transfer is slowed down because of the waves and the relevant parameter of the problem is the Rossby number, i.e. the ratio of time-scales of the wave and the nonlinear turn-over time.} Thus, we can assume that the flux will be slowed down by a factor proportional to the ratio $\tau_\Omega/\tau$ where $\tau$ is the relevant (and unknown) turnover time for the problem,
 \begin{equation}
\Pi_{2D} \sim \frac{u_\perp^2}{\tau} \frac{\tau_\Omega}{\tau} . \label{PI_2D}
\end{equation}
It is interesting to point out that if the turnover time in the above expression is built upon the velocity at the forcing scale $U_f$ as $\tau \sim l_\perp/U_f$ (i.e., assuming interactions are non-local 
{in the inverse cascade range}
and that most of the energy in the slow manifold comes directly from the 3D forced modes, which is consistent with the large and negative values of $\Pi(k_\parallel=0)$ in some of the runs), Eq.~\eqref{PI_2D} results in a $\sim k_\perp^{-3}$ scaling for the energy spectrum of the 2D modes. Note that such a choice of the time-scale is consistent with the non-local transfers reported in \cite{Bourouiba2011},
{and that non locality of nonlinear transfer in rotating turbulence in the direct cascade was also observed in \cite{Mininni09}.}

\subsection{Large-scale shear}

\begin{figure}
\begin{center}
\includegraphics[trim=0cm 6cm 0cm 7cm,clip=true,height=5.2cm,width=9.5cm]{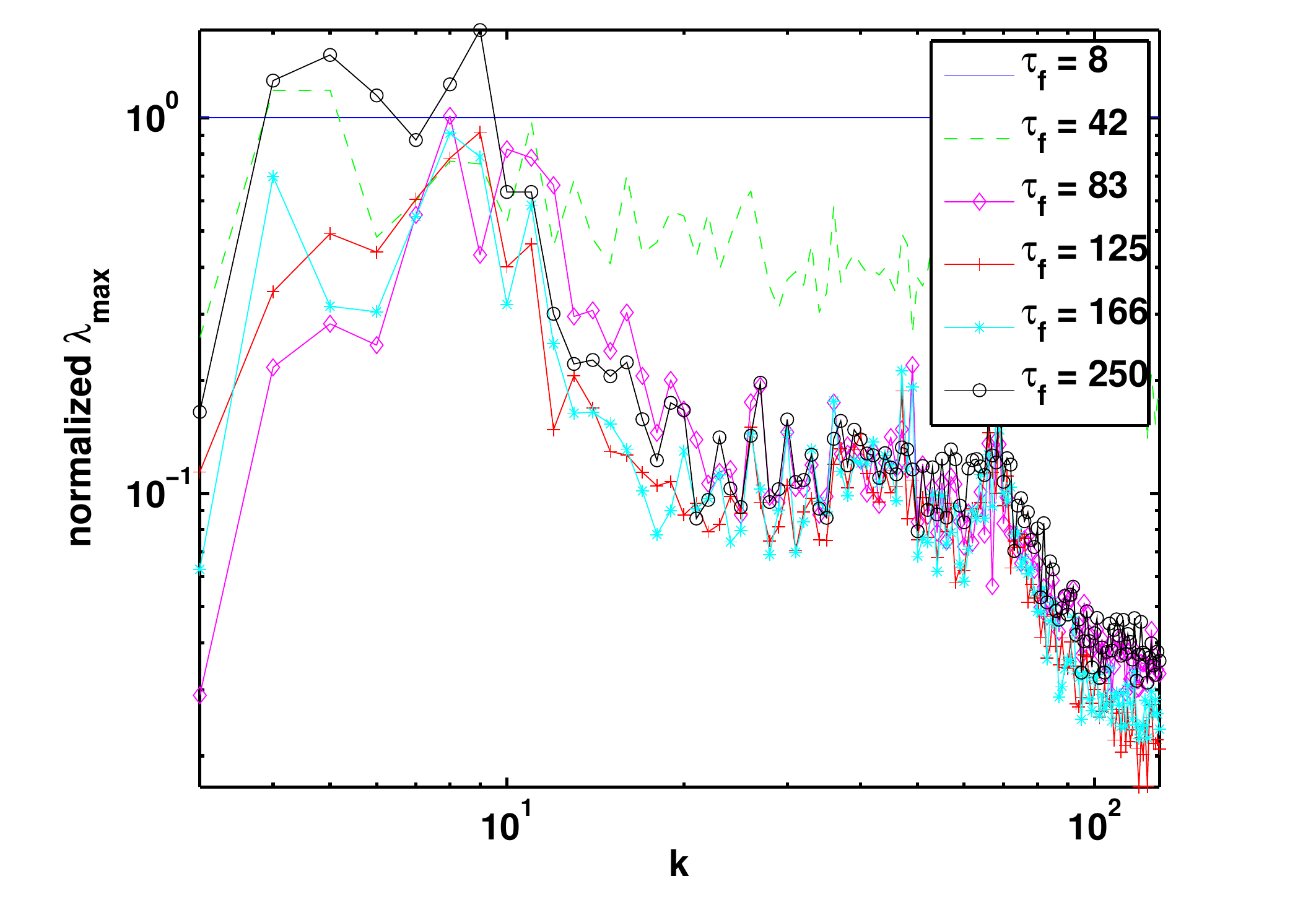}
\includegraphics[trim=0cm 6cm 0cm 7cm,clip=true,height=5.2cm,width=9.5cm]{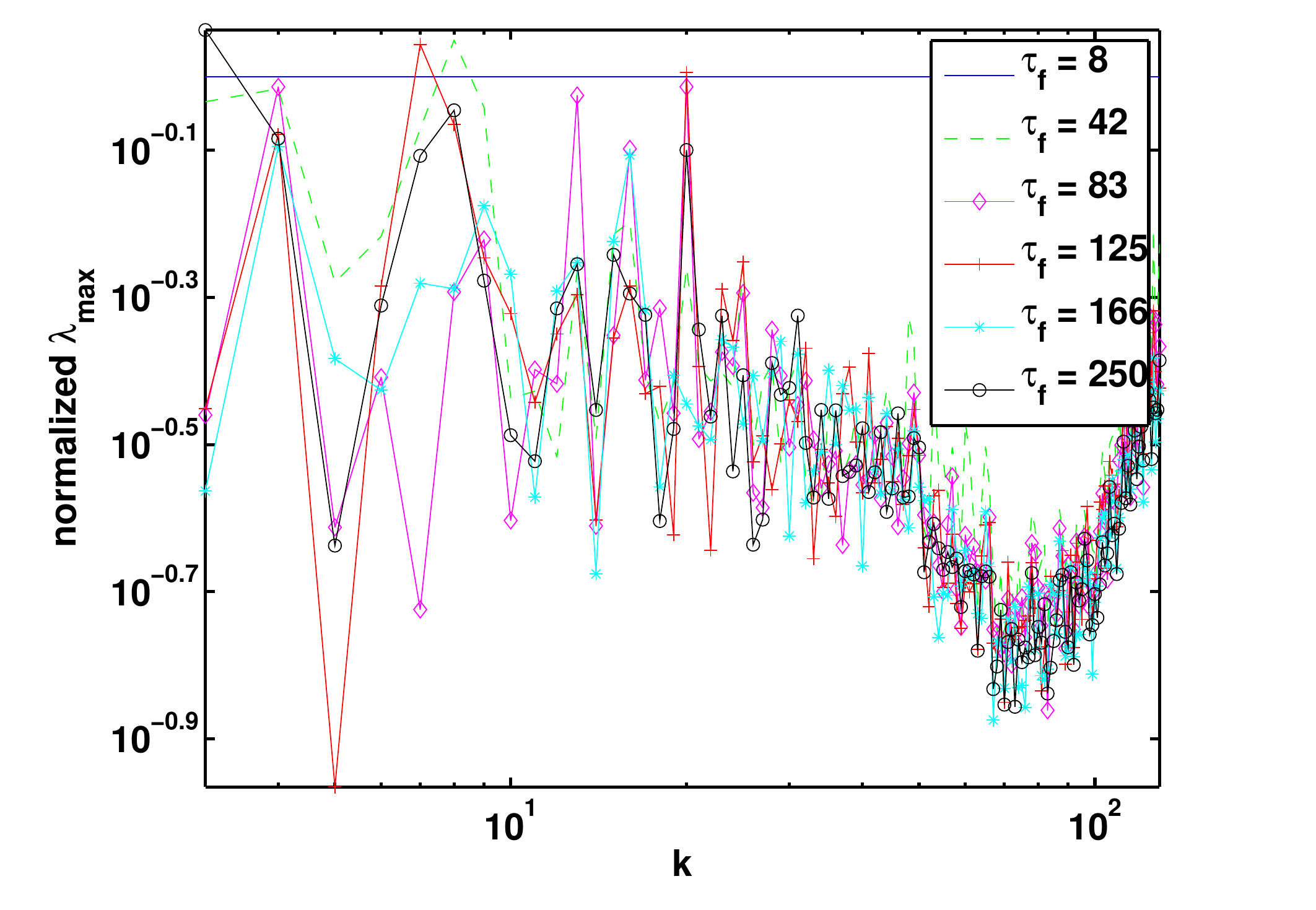}
\includegraphics[trim=0cm 6cm 0cm 7cm,clip=true,height=5.2cm,width=9.5cm]{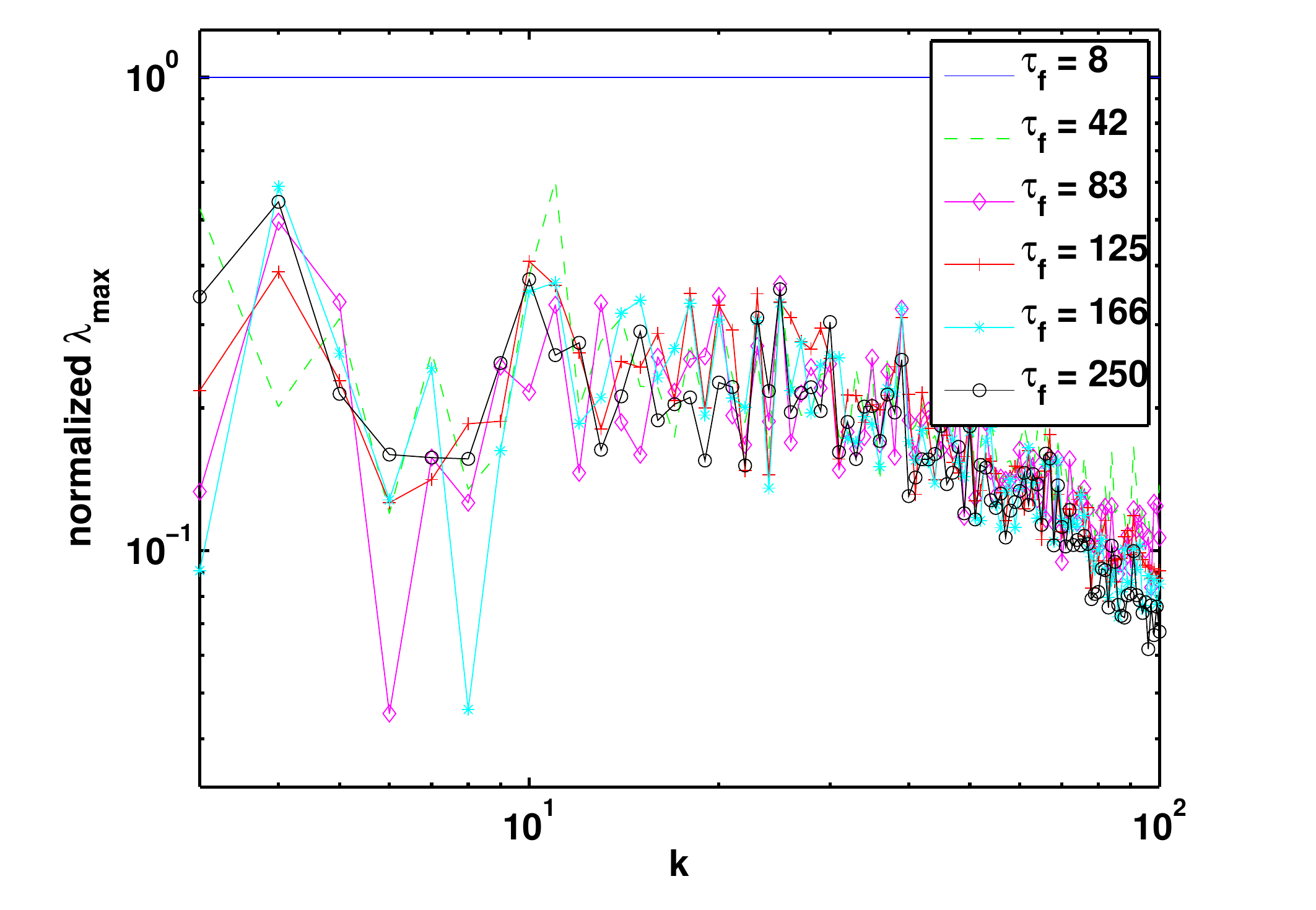}
\includegraphics[trim=0cm 6cm 0cm 7cm,clip=true,height=5.2cm,width=9.5cm]{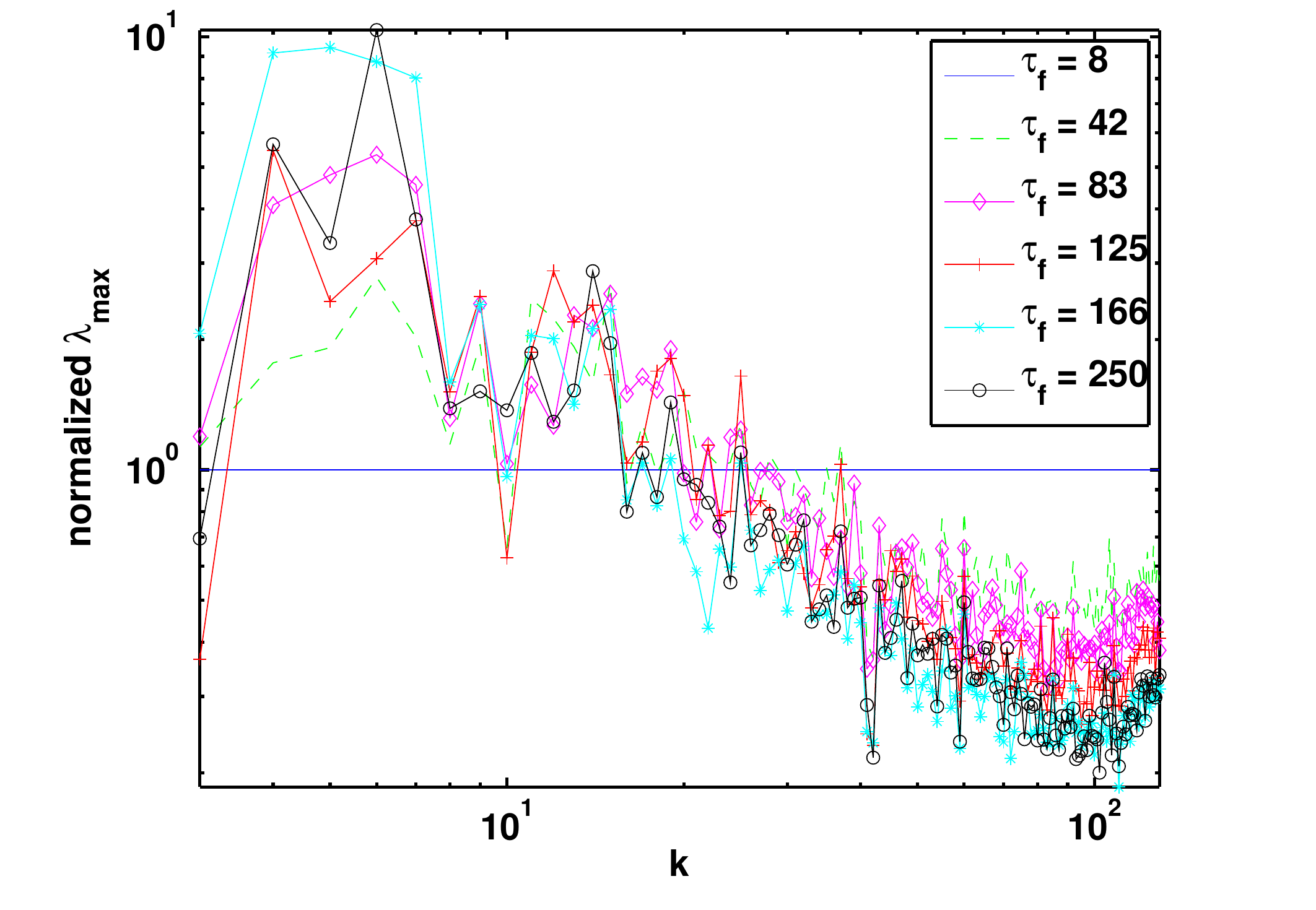}
\end{center}
\caption{Spectrum of the maximum eigenvalue of the rate of strain tensor in horizontal planes, for different runs as a function of time. The amplitudes of the spectra are normalized to their initial values (when the rotation is turned on corresponding to $\tau_f = 8$). From top to bottom: TG, RND1, RND4, and ABC.}
\label{fig:shear}
\end{figure}

The transfer of large-scale energy from the 2D modes to the 3D modes in the ABC run should have an impact in the large scale structures that develop as a result of the inverse cascade. In this section, we show that large-scales in the ABC run have large shear, while in the other runs (in which energy is mostly transferred from the 3D to the 2D modes or cases where the transfer is negligible) have much smaller shear at large scales (see Fig.~\ref{fig:shear}). The development of large-scale shear in the ABC run also introduces a new timescale for the 3D modes, $\tau_{sh}$ which is independent of the scale and therefore consistent with the observed $\sim k^{-1}$ scaling in the total energy spectrum.

To study the effect of shear, we analyze the velocity gradient tensor, $\mathfrak{V}$ defined as follows:
\begin{equation}
 \mathfrak{V} := \nabla{\bf u} = \begin{pmatrix}
                                     \partial_x u & \partial_y u & \partial_z u\\
				     \partial_x v & \partial_y v & \partial_z v\\
				     \partial_x w & \partial_y w & \partial_z w  
                                    \end{pmatrix} \ .
\end{equation}

The velocity gradient tensor may be written as the sum of the symmetric \textit{rate of strain tensor}, $\mathfrak{S}$ and the antisymmetric \textit{rotation tensor}, $\mathfrak{R}$, i.e. $\mathfrak{V} = \mathfrak{S} + \mathfrak{R}$. Note that $\mathfrak{S}$ may also be written as:
\begin{equation}
\mathfrak{S} = \frac{1}{2}(\mathfrak{V} + \mathfrak{V}^{\intercal})  \ . \label{strainTen}
\end{equation}

We analyze the spectrum of the maximum eigenvalue $\lambda_{max}$ of the rate of strain tensor $\mathfrak{S}$ in a horizontal plane for several runs (in Fig.~\ref{fig:shear}, note the amplitude of the spectra of $\lambda_{max}$ in the figure is normalized by its initial value before rotation is turned on). In most of the runs, shear decreases when rotation is turned on and the spectrum seems to reach a steady state (with significantly less shear than the isotropic and homogeneous turbulent flow) at late times. However, in the ABC run, shear increases at large scales as time evolves (with a decrease in shear at small scales). The increase in large scale shear in this run can be understood in the light of the previous discussion and based on the fact that $\Pi(k_\parallel=0)>0$. In the case of ABC forcing, $2/3$ of the energy injection corresponds to the 2D modes and the energy in the slow manifold undergoes an inverse cascade. However, some of the large-scale energy in the 2D modes is transferred back to the 3D modes (as is evident from $\Pi(k_\parallel=0)>0$)
and the excitation of large-scale 3D modes in turn creates large-scale shear (note: $\partial_z (\cdot) = 0$ in the slow manifold when $k_{\parallel}=0$ but $\partial_z(\cdot) \ne 0$ in the 3D modes).

It is interesting that once large-scale shear is present, a new constant time scale (i.e., independent of length scale) appears. Large scale shear is associated with the \textit{shear time scale}, $\tau_{sh}:= \frac{1}{\text{max}\{\lambda_{max}\}}$, where $\lambda_{max}(x,y,z)$ is the largest eigenvalue (in magnitude) of $\mathfrak{S}$ at any given location $(x,y,z)$. Dimensional analysis then hints at a {flatter} energy spectrum, and a $k^{-1}$ energy spectrum (as observed for the total energy in the ABC run) is not uncommon in shear dominated flows \cite{Perry86}. 

\section{Concluding remarks\label{sec:conclusion}}

An attempt has been made to provide a coherent treatment of the various large scale physical processes involved in rotating flows with a special emphasis on the breaking of universality in the inverse cascade of energy for rotating turbulence due to the anisotropy in the forcing.

We observe a $\sim k_{\perp}^{-3}$ spectrum in the 2D modes when the forcing is isotropic and the observed spectrum can be associated with an inverse transfer of energy. This spectrum has been reported before \cite{Bourouiba2011,Smith96} although it must be said that in \cite{Bourouiba2011}, the observed $k_{\perp}^{-3}$ spectrum is attributed to a direct enstrophy cascade, as is thought to happen in the case of two-dimensional Navier-Stokes turbulence (see \cite{trieste_12} and references therein). 

We also find that in the case of strongly anisotropic forcing such as the ABC, the inverse cascade for the 2D energy follows a $k_{\perp}^{-5/3}$ spectral law but the 3D energy follows a shallower law, $E(k)\sim k^{-1}$ attributed to the creation of shear at large scales. Of course, a $k^{-1}$ spectrum is not unheard of in turbulence; it arises for a field which is advected  at a constant rate, e.g., for a passive tracer in a turbulent flow and it is also documented in shear flows \cite{nickel} (see also the recent review in \cite{smits}). Although a recent analysis of atmospheric data indicates that the energy cascade is forward at all scales (see \cite{tellus} and references therein), it should be noted that the $k^{-3}$ and $k^{-5/3}$ spectra are both observed. 

There is no indication of inverse cascade for the helicity in any of the runs. However, it is not clear what the origin is of the rather flatter spectrum of the helicity at large scales (see Fig. \ref{fig:helispec} and also \cite{trieste_12}, Fig.~3). It is consistent with the fact that the relative helicity becomes negligible at large scales, so this could be simply interpreted as eddy-noise. This point will have to be investigated further. 
Note  that a  $k^{-1}$ scaling law, which implies less energy at large scales, may be consistent with toned down 3D non-linear dynamics due to the strong helicity in the flow.

Finally, it must be said that these results are obtained with a sub-grid scale model and due to the non-local interactions that seem to be present in the inverse cascade, the modeling of the small-scale may affect the behavior of the large scales. This effect is perhaps of less consequence when compared with models using an hyper-viscous term with the dissipation proportional to $\sim k^{2\alpha}E(k)$. Indeed, in our LES the computations of the eddy viscosity and eddy noise take into account the energy and the helicity spectra up to $3k_c$, where $k_c$ is the cut-off wavenumber. We would like to mention that DNS of the flows presented in this paper are being performed currently to substantiate the observations based on the LES.

\section{Acknowledgements}
The first author would like to acknowledge the various speakers and participants of the summer school, ANISO 2011 on \textit{Morphology and dynamics of anisotropic flows} held at the Institute of Theoretical Physics, Carg\`ese, Corsica for useful discussions. This work was sponsored by an NSF cooperative agreement through the University Corporation for Atmospheric Research on behalf of the National Center for Atmospheric Research (NCAR). Computer time was provided by NSF under sponsorship of NCAR and by way of NSF TeraGrid (project numbers ASC090050, TG-PHY100029).
{NSF/CMG grant 1025183 is gratefully acknowledged.}

\end{document}